\documentclass[12pt]{article}
\usepackage{color}
\usepackage{graphicx}
\usepackage{slashed}
\usepackage{geometry}
\begin{document}
\title{Spectroscopy of ground and excited states of pseudoscalar and vector charmonium and bottomonium}
\author{Hluf Negash and Shashank Bhatnagar}
\maketitle
Department of Physics, Addis Ababa University, P.O.Box 1176, Addis Ababa, Ethiopia\\

\textbf{Abstract}\\
In this work we calculate the mass spectrum, weak decay constants,
two photon decay widths, and two gluon decay widths of ground
(1S), and radially excited (2S, 3S,...) states of pseudoscalar
charmoniuum and bottomonium such as $\eta_{c}$ and $\eta_{b}$, as
well as the mass spectrum and leptonic decay constants of ground
state (1S), excited (2S, 1D, 3S, 2D, 4S,...,5D) states of vector
charmonium and bottomonium such as $J/\psi$, and $\Upsilon$, using
the formulation of Bethe-Salpeter equation under covariant
Instantaneous Ansatz (CIA).  Our results are in good agreement
with data (where ever available) and other models. In this
framework, from the beginning, we employ a $4\times 4$
representation for two-body ($q\overline{q}$) BS amplitude for
calculating both the mass spectra as well as the transition
amplitudes. However, the price we have to pay is to solve a
coupled set of equations for both pseudoscalar and vector
quarkonia, which we have explicitly shown get decoupled in the
heavy-quark approximation, leading to mass spectral equation with
analytical solutions for both masses, as well as eigenfunctions
for all the above states, in an approximate harmonic oscillator
basis. The analytical forms of eigen functions for ground and
excited states so obtained are used to evaluate the decay
constants and decay widths for different processes.

\section{Introduction}

Heavy flavor mesons are of the type $Q\overline{Q}$, or
$Q\overline{q}$, in which one of the quarks belongs to a heavy
flavour such as $c$ or $b$. Quarkonia usually refers to heavy
mesons with quark composition $(Q\overline{Q})$. Charmonium and
bottomonium are two types of quarkonia that are observed with
quark composition $c\overline{c}$ and $b\overline{b}$
respectively. Physically, the charmonium system is analogous to
positronium system. However, historically, its role as a model
system in QCD is similar in importance to the role of Hydrogen
atom in Quantum mechanics, due to the fact that this system is
non-relativistic and has a hydrogen atom like spectrum. Charmonium
in fact occupies a valuable intermediate position within QCD,
being neither in the purely non-relativistic regime nor the regime
where chiral symmetry breaking dominates the physics. This makes
it a relatively clean system in which to study non-perturbative
QCD dynamics, QCD-inspired quark-potential models as well as
lattice QCD, which have been rather successful in describing the
observed features of the spectrum (see \cite{swanson06} for a
review). However, charmonium cannot be considered to be completely
understood; as an example, in recent years a number of new
charmonium resonances have been claimed in experiment, several of
which cannot be easily reconciled with the predictions of simple
quark-potential models. Charmonium spectrum provided a simple
example of how QCD works, which was made even more compelling with
the subsequent observation of bottomonium spectra. Investigation
of properties of these mesons gives a lot of information about
heavy quark dynamics. Below the heavy flavour meson  pair
($D\overline{D}$) production threshold, all charmonia states have
been well established, and there is a good agreement between
predictions based on potential models and data. However, there are
many charmonium and charmonium like states
\cite{brambilla11,G,Ecklund,Auger} observed above $D\overline{D}$
threshold in the past ten years. Some of these are good candidates
for the charmonia predicted in different models. However, many
have unusual quantum numbers \cite{brambilla11,G,Ecklund,Auger}
(which can not be reached by pure quark states), indicating that
exotic states such as multi - quark states, molecule, hybrid, or
glueball may have been observed. The spectroscopy and decay rates
of quarkonia are quite important to study as huge amount of high
precession data acquired from many experimental facilities world
over are continuously providing accurate information about hadrons
particularly in charm and beauty flavour sectors. The mass
spectrum and the decays of these bound states can be tested
experimentally and thus studies on them  may throw valuable
insight of the heavy quark dynamics and lead to a deeper
understanding of QCD further.
\bigskip

As regards the dynamical framework, to investigate these
properties is concerned, many non-perturbative approaches, such as
Lattice QCD \cite{mcnielle12,bali97,burch09}, Chiral perturbation
theory \cite{gasser84}, heavy quark effective theory
\cite{neubert94}, QCD sum rules \cite{shifman79,veli12}, N.R.QCD
\cite{bodwin}, dynamical-equation based approaches like
Schwinger-Dyson equation and Bethe-Salpeter equation (BSE)
\cite{smith69,mitra01,munz,alkofer01,wang10,koll,bhatnagar92}, and
potential models \cite{bhagyesh11, godfrey85} have been proposed
to deal with the long distance property of QCD . Some of the
interesting works on heavy-quark spectroscopy in the framework of
BSE are \cite{babutsidze,yang95}.

Many theoretical predictions on the decay properties particularly
the leptonic and two-gamma decays of quarkonia based on the
relativistic quark model or potential model
\cite{Ahmady,ebert,Hwang}, Bethe-Salpeter
equation\cite{Huang,kim}, heavy-quark spin symmetry
\cite{lansberg} and lattice QCD \cite{JJ} are available in
literature. Such studies have become a hot topic in recent years,
due to observation of many new states at various high energy
accelerators at BABAR, Belle, CLEO and BES-III collaborations
\cite{babar09,belle10,cleo01,olive14}. All this has opened up new
challenges in theoretical understanding of heavy hadrons and
provide an important tool for exploring the structure of these
simplest bound states in QCD and for studying the non-perturbative
(long distance) behavior of strong interactions.
\bigskip

We wish to mention that though heavy quarkonia ($c\overline{c}$
and $b\overline{b}$) are very well described in terms of NR
potential models, and relativized potential models (like
\cite{godfrey85}) can treat light mesons simultaneously with heavy
mesons, however, they cannot be regarded as an ideal theory that
is closest to QCD, due to their inability in implementing Lorentz
and gauge invariance for a consistent treatment of motion of
quarks within the hadron.  Most of  such models, employ  the
$v^{2}/c^{2}$ corrections, or the replacement
$\frac{p^{2}}{2m}\rightarrow (p^{2}+ m^{2})^{1/2}$ in the kinetic
energy operator (for each quark) in a Schrodinger type equation,
which provides the bulk of momentum dependent effects. However,
interaction terms require a more substantial form of treatment
consistent with relativity. We further wish to point out that
though, in heavy quarkonia, NR treatment may at best be a good
first approximation,  but  in charmonium (where charmed quark is
not heavy enough to be considered non-relativistically), one finds
typical velocities, $\frac{v}{c}\sim 0.4$, making relativistic
effects in charmonium important, which is specially so for
electroweak decay properties as well as various other transitions
at high energies, as mentioned above. These decay widths are
sensitive to relativistic effects (see \cite{munz}). Thus NR and
semi-relativistic equations (as in \cite{godfrey85}) are not
completely adequate as proper starting point for light, and
semi-light  quark systems.

In this respect, the Bethe-Salpeter Equation (BSE), with its 4D
Lorentz-invariant structure is a natural candidate for hadron
physics, since it is firmly rooted in field theory, and its
general framework rests on Lorentz and gauge invariance, due to
which it has a very wide applicability, all the way from low
energy spectroscopy to high energy transition amplitudes, and can
be used to study both light and heavy quark systems in an
integrated manner. Thus the BSE framework provides a realistic
description for analyzing hadrons as composite objects. Despite
its drawback of having to input model-dependent kernel, these
studies have become an interesting topic in recent years, since
calculations have shown that BSE framework using phenomenological
potentials can give satisfactory results on more and more data are
being accumulated. The Lorentz-invariance at the input form of BSE
formally ensures a continued validity of predictive powers from
low energy spectroscopy to high energy processes within a common
dynamical framework. The 4D BS wave function
\cite{mitra01,mitra99,bhatnagar91,bhatnagar06,bhatnagar11,bhatnagar14}
(that carries the entire non-perturbative information) arises as
solution of the 4D BS equation.  Since this wave function is
determined as a solution of the formal dynamical equation, it does
not suffer from the uncertainties of a variational determination
(as in NR approaches \cite{godfrey85}), and is able to carry over
the micro causality information contained in the BSE.  From
non-perturbative 4D BS wave function, one can work out the 4D
hadron-quark vertices, that are effective coupling vertices of
hadron with all its constituents (quarks), and then naturally work
out transition amplitudes for different processes over a vast
range of energies, employing Feynman diagrammatic techniques over
various quark-loop diagrams corresponding to the process
considered.

\bigskip

A number of recent studies in the framework of BSE have been
carried out on leptonic decays and two photon decays of mesons.
The decay into two photons is considered as an interesting
experimental playground in the mesonic physics of the near future.
The two-photon decay of mesons can be used to identify the flavor
of quark and anti-quark states. Two photon couplings also provide
a useful probe of the internal structure of mesons. The states
such as, $\eta_{c}$ and $\eta_{b}$ have been observed to have
partial widths consistent with quark model predictions. These
states can also decay into two gluons \cite{laverty}, which
accounts for a substantial portion of the hadronic decays for
states below the $c\overline{c}$ or $b\overline{b}$ threshold. It
is clearly important to have accurate quark model predictions of
widths for all experimentally accessible $q\bar{q}$-mesons. Thus,
in this work we are predominantly interested in studying the mass
spectrum and leptonic decays of ground and excited states of
pseudoscalar and vector quarkonia, such as $\eta_c$, $\eta_b$,
 $J/\psi$, and $\Upsilon$ respectively (which proceed through the
coupling of quark- anti-quark loop to the axial vector and vector
currents respectively), as well as, the two-photon radiative
decays, and two-gluon decays of ground state ($1S$) and radial
excitations ($2S$ and $3S$) of conventional pseudoscalar mesons
such as $\eta_c$ and $\eta_b$ in the framework of BSE under
Covariant Instantaneous Ansatz (CIA)
\cite{mitra01,bhatnagar92,bhatnagar91,bhatnagar06,bhatnagar11,bhatnagar14},
though some of the excited states have not yet been experimentally
discovered.
\bigskip

We have employed the BSE framework under Covariant Instantaneous
Ansatz (CIA). CIA is a Lorentz-invariant generalization of the
instantaneous ansatz (IA). What distinguishes CIA from other
three-dimensional (3D) reductions of BSE is its capacity for a
two-way interconnection
\cite{mitra01,bhatnagar92,bhatnagar91,bhatnagar14}: an exact 3D
BSE reduction for a $q\overline{q}$ system (for calculation of the
mass spectrum), and an equally exact reconstruction of original 4D
BSE (for calculation of transition amplitudes as 4D quark loop
integrals). In some of our previous works in BSE under CIA, we had
made use of a confinement potential $V(\widehat{q},\widehat{q}')$
given in Eq.(33), for a number of successful predictions for not
only the mass spectrum \cite{mitra01,mitra99,bhatnagar14} of light
$(q\overline{q})$ and some heavy mesons, but also a number of
processes such as leptonic decays of vector mesons such as: $\rho,
\omega,\phi,J/\psi$, and $\Upsilon$ \cite{bhatnagar14} through the
process $V\rightarrow e^{+}+e^{-}$, leptonic decays of
pseudoscalar mesons such as $\pi, K, D, D_{S}$ and $B$ through the
process, $P\rightarrow l+\overline{\nu_{\overline{l}}}$  and
two-photon decays of $\pi$, and $\eta_{c}$ mesons
\cite{bhatnagar11} through process, $P\rightarrow \gamma+\gamma$,
as well as the processes proceeding through quark triangle loop
diagrams involving two or more hadron-quark vertices, such as:
single photon decays of $\rho^{0,\pm}$, and $\omega^{0}$ mesons
through process $V\rightarrow P+\gamma$, as well as strong decays
of first radial excitations of lightest vector meson states
$\rho(1450)$, and $\omega(1420)$\cite{bhatnagar13} through process
$V'\rightarrow V+P$ (where $V$, and $P$ refer to vector and
pseudoscalar mesons). However this form of confining potential was
employed mostly for light mesons, and for some cases to heavy
mesons. However, in our previous BSE framework
\cite{bhatnagar06,bhatnagar11,bhatnagar14}, we had adopted a
$16\times 1$ column representation for two-body ($q\overline{q}$
or $qq$) BS amplitude (though both $16\times 1$ and $4\times 4$
representations of BSE are completely equivalent \cite{smith69}).
We had employed Gordon reduction on the mass shells of individual
quarks to obtain the 3D mass spectral equation. And for bringing
out the structure of the full 4D BS wave function in a $4\times 4$
matrix form (that is needed for calculation of transition
amplitudes through quark-loop diagrams), we made use of standard
transformations \cite{smith69} of charge conjugation of spinors.
This is in contrast to the BSE framework employed in this paper,
where we have employed a $4\times 4$ framework of BSE from the
very beginning (with the same confining potential in Eq.(33)), and
used it for both the mass spectral predictions as well as the
transition amplitude calculations involving heavy quarkonia (both
pseudoscalar and vector), though the price one has to pay for this
approach, is to solve a set of coupled equations, to obtain the
masss spectral equations. However, in the present work, we have
explicitly shown that in the heavy quark approximation (valid for
$c\overline{c}$, and $b\overline{b}$ systems), these equations can
be decoupled, and analytical solutions (both mass spectrum and
eigen functions) of these equations can be obtained using
approximate harmonic oscillator basis.
\bigskip

\begin{table}[h]
  \begin{center}
\begin{tabular}{lllllllllll}
\hline
     $L^{2S+1}_{J}$ &$S^{1}_{0}$&$S^{3}_{1}$&$P^{1}_{1}$&$P^{3}_{0}$&$P^{3}_{1}$&$P^{3}_{2}$&$D^{1}_{2}$&$D^{3}_{1}$&$D^{3}_{2}$&$D^{3}_{3}$\\ \hline
    $J^{PC}$&$0^{-+}$&$1^{--}$&$1^{+-}$&$0^{++}$&$1^{++}$&$2^{++}$&$2^{-+}$&$1^{--}$&$2^{--}$&$3^{--}$\\
    \hline
      \end{tabular}
      \end{center}
  \caption{Quantum numbers of lowest quarkonium states. These quantum numbers are repeated for the radial excitations of these states}
    \end{table}

\bigskip

Towards this end, we first analyze the quarkonium states according
to their total angular momentum,
$\overrightarrow{J}=\overrightarrow{L}+\overrightarrow{S}$, parity
$P=(-1)^{L+1}$, and charge conjugation, $C=(-1)^{L+S}$ for states
classified as $J^{PC}$, with the lowest possible states for
quarkonia are listed in Table 1. We start with the most general
structure of Bethe-Salpeter wave function for $J^{PC}=0^{-+}$ and
$J^{PC}=1^{--}$. It can be seen from Table 1, that $1^{--}$ state
has not only the ground state ($1S$) components, but also the
orbital ($1D$) excitations. The same holds true for their radial
excitations. We have started with the full Dirac structure of
these states with all the Dirac covariants multiplying various
scalar functions of internal hadron momentum, $q$ as in
\cite{smith69}, with various Dirac structures incorporated into
the wave functions in accordance with the power counting rule
\cite{bhatnagar06,bhatnagar14} suggested recently. We then put the
formulated BS wave functions into the instantaneous BSE and turn
the equation into a set of proper coupled equations \cite{hluf15}
for the components which appear in the formulation. These
equations are then explicitly shown to decouple in the heavy-quark
limit, and are reduced to a single mass spectral equation, whose
analytic solutions in an approximate harmonic oscillator basis
yield the mass spectrum and wave functions for ground and excited
states of $\eta_c$, $\eta_b$, $J/\psi$ and $\Upsilon$. For this we
need 6 input parameters (that include two input quarks masses
$m_c$, and $m_b$). We then derive the leptonic decay constants,
and the decay widths for two-photon decays, and two-gluon decays
of pseudoscalar quarkonia $\eta_c$, and $\eta_b$ for their ground
and radially excited states, as well as the leptonic decay
constants for ground, radially and orbitally excited states of
vector quarkonia, $J/\psi$, and $\Upsilon$, with the 6 input
parameters fixed above.

\bigskip
This paper is structured as follows: in section \textbf{2} we
introduce the BSE and formulate the instantaneous BSE for
pseudoscalar and vector quarkonia. In section \textbf{3} we start
with the generalized formulation of BS wave functions for
pseudoscalar ($J^{PC}=0^{-+}$) and vector ($J^{PC}=1^{--}$)
quarkonia, with definite quantum numbers and derive their mass
spectral coupled equations. In section \textbf{4}, we study the
leptonic decays of pseudoscalar and vector quarkonia. In section
\textbf{5}, we derive the two photon and two gluon decays of
pseudoscalar quarkonia. Finally, section \textbf{6} is relegated to
numerical results and discussion.

\section{Formulation of BSE under CIA}
\bigskip
Lets consider a $q\bar{q}$ comprising of fermionic quarks of masses
$m_{1}$ and $m_{2}$ respectively. We start with a 4D BSE for
$q\bar{q}$ system, written in a $4\times 4$ representation of 4D BS
wave function $\Psi(P,q)$ as:

\begin{equation}
S_{F}^{-1}(p_{1})\Psi(P,q)S_{F}^{-1}(-p_{2}) =
\frac{i}{(2\pi)^{4}}\int d^{4}q'K(q,q')\Psi(P,q')
\end{equation}

where $K(q,q')$ is the interaction kernel between the quark and
anti-quark, and $p_{1,2}$ are the momenta of the quark and
anti-quark, which are related to the internal 4-momentum $q$ and
total momentum $P$ of hadron of mass $M$ as,

\begin{equation}
p_{1,2\mu} = \hat{m}_{1,2}P_{\mu} \pm q_{\mu}
\end{equation}

where
$\hat{m}_{1,2}=\frac{1}{2}[1\pm\frac{(m^{2}_{1}-m^{2}_{2})}{M^{2}}]$
are the Wightman-Garding (WG) definitions of masses of individual
quarks which ensure that on the mass shells ($P.q=0$) of either
quarks, even when $m_1\neq m_2$. However for equal mass mesons,
($m_{1}=m_{2}=m$), we have $\hat{m}_{1}=\hat{m}_{2}=\frac{1}{2}$.

Then $p_{1,2\mu}$ becomes,
\begin{equation}
p_{1,2\mu} = \frac{1}{2}P_{\mu} \pm q_{\mu}
\end{equation}

Now it is convenient to express the internal momentum of the
hadron $q_\mu$ as the sum of two parts. They are: (i) the
transverse component, $\hat{q}_\mu=q_\mu-(q\cdot P)P_\mu/P^2$
which is orthogonal to total hadron momentum $P_\mu$ (ie.
$\widehat{q}\cdot P=0$ regardless of whether the individual quarks
are on-shell or off-shell), and (ii) the longitudinal component,
$\sigma P_\mu = (q\cdot P)P_\mu/P^2$, which is parallel to
$P_\mu$. Thus we can decompose $q_\mu$ as,
$q_\mu=(M\sigma,\widehat{q})$, where the transverse component,
$\widehat{q}$ is an effective 3D vector, while the longitudinal
component, $M\sigma$ plays the role of the time component. The 4-D
volume element in this decomposition is,
$d^4q=d^3\hat{q}Md\sigma$. To obtain the 3D BSE and the
hadron-quark vertex, use an Ansatz on the BS kernel $K$ in Eq. (1)
which is assumed to depend on the 3D variables $\hat{q}_\mu$,
$\hat{q}_\mu^\prime$ as,
\begin{equation}
 K(q,q') = K(\hat{q},\hat{q}')
\end{equation}
Hence, the longitudinal component, $M\sigma$ of $q_{\mu}$, does not
appear in the form $K(\hat{q},\hat{q}')$ of the kernel. For reducing
Eq.(1) to 3D form, we define 3D wave function $\psi(\hat{q})$ as:
\begin{equation}
\psi(\hat{q}) = \frac{i}{2\pi}\int Md\sigma\Psi(P,q)
\end{equation}
Substituting Eq.(5) in eq.(1), with definition of kernel in eq.(4),
we get a covariant version of Salpeter equation,
\begin{equation}
\ (\slashed{p}_{1}-m_{1})\Psi(P,q)(\slashed{p}_{2}+m_{2}) = \int
\frac{d^{3}\hat{q}'}{(2\pi)^{3}}
K(\hat{q},\hat{q}')\psi(\hat{q}'),
\end{equation}

and the 4D BS wave function becomes,
\begin{equation}
\Psi(P,q) = S_{F}(p_{1})\Gamma(\hat{q})S_{F}(-p_{2})
\end{equation}
where
\begin{equation}
\Gamma(\hat{q})=\int \frac{d^{3}\hat{q}'}{(2\pi)^{3}}
K(\hat{q},\hat{q}')\psi(\hat{q}')
\end{equation}
 plays the role of hadron-quark vertex function. The kernel is taken to have $\gamma_{\mu}\otimes\gamma^{\mu}$ form,
  details of which
 are given in the next section. Thus with this form of the kernel,
 we can write $\Gamma(\hat{q})=\int \frac{d^{3}\hat{q}'}{(2\pi)^{3}}
V(\hat{q},\hat{q}')\gamma_{\mu}\psi(\hat{q}')\gamma^{\mu}$ (where
$V$ is spatial part of the kernel). We can to a good approximation
express $\gamma_{\mu}\psi(\hat{q}')\gamma^{\mu}\approx
\Theta\psi(\hat{q}')$, where $\Theta$ involves the spin-spin
interactions alone, that factor out of the RHS of the hadron-quark
vertex on taking the dominant Dirac structures in $\psi(\hat{q}')$
in the calculation of $\Theta$, and we can write
$\Gamma(\hat{q})=\Theta\int \frac{d^{3}\hat{q}'}{(2\pi)^{3}}
V(\hat{q},\hat{q}')\psi(\hat{q}')$. And, $S_{F}(p)$ is the
 usual fermionic propagator
of the quarks, given as,
\begin{equation}
\ S_{F}(\pm p_{1,2}) = \frac{\slashed{p}_{1,2} \pm
m_{1,2}}{\Delta_{1,2}}
\end{equation}
where $\Delta_{1,2}=p_{1,2}^{2}\mp m_{1,2}^{2}$, which can also be
decomposed as in\cite{koll} ,
\begin{equation}
S_{F}(\pm p_{i})=
\frac{\Lambda_{i}^{+}(\hat{q})}{I(i)M\sigma+\frac{1}{2}M-\omega_{i}}
    +\frac{\Lambda_{i}^{-}(\hat{q})}{I(i)M\sigma+\frac{1}{2}M+\omega_{i}}
\end{equation}
with
\begin{eqnarray}
 &&\nonumber\omega_{i}^{2} = m_{i}^{2} + \hat{q}^{2}\\&&
 \Lambda_{i}^{\pm}(\hat{q})=\frac{1}{2\omega_{i}}[\frac{\slashed{P}}{M}\omega_{i}\pm
 I(i)(m_{i}+\slashed{\hat{q}})
 ]
\end{eqnarray}
where $i=1,2$ for quark and anti-quark respectively, and
$I(i)=(-1)^{i+1}$. Here $\Lambda_{i}^{\pm}(\hat{q})$, are called
as projection operators. With the projected wave functions, one
can rewrite the BSE as,
 \begin{equation}
(\frac{1}{2}M\mp\omega_{1}+M\sigma)(\frac{1}{2}M\mp\omega_{2}-M\sigma)\Psi^{\pm\pm}(P,q)
=
\Lambda_{1}^{\pm}(\hat{q})\Gamma(\hat{q})\Lambda_{2}^{\pm}(\hat{q}),
\end{equation}
where projected wave functions, $\psi^{\pm\pm}(\hat{q})$ are
obtained by the operation of projection operators on
$\psi(\widehat{q})$ as,
\begin{equation}
 \psi^{\pm\pm}(\hat{q})=\Lambda_{1}^{\pm}(\hat{q})\frac{\slashed{P}}{M}\psi(\hat{q})
 \frac{\slashed{P}}{M}\Lambda_{2}^{\pm}(\hat{q}).
\end{equation}
With contour integration over $d\sigma$ on both sides of Eq.(12), we
can obtain:
\begin{equation}
 \psi(\hat{q})=-\frac{\Lambda_{1}^{+}(\hat{q})\Gamma(\hat{q})\Lambda_{2}^{+}(\hat{q})}{M-\omega_{1}-\omega_{2}}+
 \frac{\Lambda_{1}^{-}(\hat{q})\Gamma(\hat{q})\Lambda_{2}^{-}(\hat{q})}{M+\omega_{1}+\omega_{2}}
\end{equation}

The complete wave function can separate in four parts as:
\begin{equation}
 \psi(\hat{q})=\psi^{++}(\hat{q})+\psi^{+-}(\hat{q})+\psi^{-+}(\hat{q})+\psi^{--}(\hat{q}).
\end{equation}
The BSE then reduces to four independent equations as \cite{hluf15}:
\begin{eqnarray}
 &&\nonumber(M-2\omega)\psi^{++}(\hat{q})=-\Lambda_{1}^{+}(\hat{q})\Gamma(\hat{q})\Lambda_{2}^{+}(\hat{q})\\&&
   \nonumber(M+2\omega)\psi^{--}(\hat{q})=\Lambda_{1}^{-}(\hat{q})\Gamma(\hat{q})\Lambda_{2}^{-}(\hat{q})\\&&
 \psi^{+-}(\hat{q})=\psi^{-+}(\hat{q})=0
\end{eqnarray}
where $\omega_{1}=\omega_{2}(=\omega)$, for equal mass systems. In
fact the four equations constitute an eigenvalue problem that
should lead to evaluation of mass spectra of pseudoscalar (for
preliminary work, see \cite{hluf15}) and vector charmonium and
bottomonium states such as $\eta_{c}$, $\eta_{b}$, $J/\psi$ and
$\Upsilon$. The framework is quite general so far. Thus to obtain
the mass spectral equation,
we have to start with the above four equations to solve the instantaneous BS equation.\\

\section{Derivation of mass spectral equations of pseudoscalar and vector quarkonia}

\bigskip
In this section, we show the method to solve an Instantaneous BS
equation, which is first applied to the calculation of the mass
spectrum of equal mass heavy pseudoscalar and vector quarkonia. We
first write down the most general formulation of the relativistic
BS wave functions, according to the total angular momentum (J),
parity (P) and charge conjugation(C) of the concerned bound state.
We then put this BS wave function in Eq.(16) to derive the mass
spectral equations, which are a set of coupled equations. We now
illustrate this procedure for pseudoscalar and vector quarkonia.

\begin{itemize}
\item For pseudoscalar mesons, the complete decomposition of 4D BS wave function in terms of
various Dirac structures and scalar functions $\phi_{j}(P,q)$
multiplying them is \cite{smith69}:

\begin{equation}
\Psi^{P}(P,q) =
\{\phi_{1}(P,q)+\slashed{P}\phi_{2}(P,q)+\slashed{q}\phi_{3}(P,q)
+[\slashed{P},\slashed{q}]\phi_{4}(P,q)\}\gamma_{5}
\end{equation}
where $\phi_{j}=\phi_{j}(q^{2},q.P,P^{2}); (j=1,2,3,4)$ are the
Lorentz scalar amplitudes multiplying the various Dirac structures
in the BS wave function, $\Psi(P,q)$.
 Now in our framework, by use of a naive power counting rule in \cite{bhatnagar06,bhatnagar11,bhatnagar14},
we had shown that the Dirac structures associated with amplitudes
$\phi_{1}$ and $\phi_{2}$ are leading, while the structures
associated with $\phi_{3}$ and $\phi_{4}$ are sub-leading, and
would contribute lesser to meson observable calculations in
comparison to leading Dirac structures associated with $\phi_{1}$
and $\phi_{2}$. And in various calculations
\cite{bhatnagar06,bhatnagar11,bhatnagar14}, we had shown that
among the two leading Dirac structures associated with amplitudes
$\phi_{1}$, and $\phi_{2}$, the structure associated with
$\phi_{1}$ (i.e. $\gamma_{5}$) is dominant. We wish to point out
that Munczek and Jain \cite{munczek} have also earlier shown that
$\phi_{1}$ is the dominant amplitude for all ground state
pseudoscalar mesons. V.Sauli \cite{sauli} has recently shown that
the same is valid for their excited states as well. And this is
particularly true for mesons made from heavy flavour quarks.

In the center of mass frame, where $q_{\mu}=(0,\widehat{q})$, we can
then write the general decomposition of the instantaneous BS wave
function for pseudoscalar mesons $(J^{PC}=0^{-+})$, of
dimensionality $M$ in the center of mass frame as \cite{wang10}:
\begin{equation}
\psi^{P}(\hat{q}) \approx
[M\phi_{1}(\hat{q})+\slashed{P}\phi_{2}(\hat{q})+\slashed{\hat{q}}\phi_{3}(\hat{q})
+\frac{\slashed{P}\slashed{\hat{q}}}{M}\phi_{4}(\hat{q})]\gamma_{5},
\end{equation}
where $\phi_{1}$, $\phi_{2}$, $\phi_{3}$ and $\phi_{4}$ are even
functions of $\hat{q}$ and M is the mass of the bound state (of the
corresponding meson). We now obtain the algebraic forms of these
amplitudes.

For this, we put the instantaneous BS wave function
$\psi^{P}(\widehat{q})$ (Eq.(18)) into the last two equations of
Eq.(16), and this leads to independent constraints on the
components for the Instantaneous BS wave function as in
\cite{wang10,koll,yang95}:
\begin{equation}
\phi_{3}=0 ;    \phi_{4}=\frac{-\phi_{2}M}{m}
\end{equation}
So we can apply the obtained constraints Eq.(19) to Eq.(18) and
rewrite the relativistic wave function of state $(0^{-+})$ as,
\begin{equation}
\psi^{P}(\hat{q}) \approx
[M\phi_{1}(\hat{q})+\slashed{P}\phi_{2}(\hat{q})+\frac{\slashed{\hat{q}}\slashed{P}}{m}\phi_{2}(\hat{q})]\gamma_{5}.
\end{equation}
One can then see that the Instantaneous BS wave function of
$0^{-+}$ state is determined by only two independent functions
$\phi_{1}$, and $\phi_{2}$. Putting the wave function Eq.(20) into
the first two equations of Eq.(16) and by evaluating trace over
the $\gamma$-matrices on both sides, we obtain two BS coupled
integral equations,
\begin{eqnarray}
 &&\nonumber(M-2\omega)\left[\phi_{1}(\hat{q})+\frac{\omega}{m}(1-\frac{\hat{q}^{2}}{m^{2}})\phi_{2}(\hat{q})\right]=
 \Theta_{P}\int\frac{d^{3}\hat{q}'}{(2\pi)^{3}}
 V(\hat{q},\hat{q}')\left[\phi_{1}(\hat{q}')+\frac{\omega}{m}(1+\frac{\hat{q}.\hat{q}')}{m^{2}}\phi_{2}(\hat{q}')\right]\\&&
 \ (M+2\omega)\left[\phi_{1}(\hat{q})-\frac{\omega}{m}(1-\frac{\hat{q}^{2}}{m^{3}})\phi_{2}(\hat{q})\right]=
 -\Theta_{P}\int\frac{d^{3}\hat{q}'}{(2\pi)^{3}}
 V(\hat{q},\hat{q}')\left[\phi_{1}(\hat{q}')-\frac{\omega}{m}(1+\frac{\hat{q}.\hat{q}'}{m^{3}})\phi_{2}(\hat{q}')\right].
\end{eqnarray}
To decouple these equations, we first add them. Then we subtract
the second equation from the first equation. For a kernel that can
be expressed as
$V(\widehat{q}-\widehat{q}')=\overline{V}(\widehat{q})\delta^{3}(\widehat{q}-\widehat{q}')$,
we get two algebraic equations which are still coupled. Then from
one of the two equations so obtained, we eliminate
$\phi_{1}(\hat{q})$ in terms of $\phi_{2}(\hat{q})$, and plug this
expression for $\phi_{1}(\hat{q})$ in the second equation of the
coupled set so obtained to get a decoupled equation in
$\phi_{2}(\hat{q})$. Similarly, we eliminate $\phi_{2}(\hat{q})$
from the second equation of the set of coupled algebraic equations
in terms of $\phi_{1}(\hat{q})$, and plug it into the first
equation to get a decoupled equation entirely in
$\phi_{1}(\hat{q})$. Thus, we get two identical decoupled
equations, one entirely in $\phi_{1}(\hat{q})$, and the other that
is entirely in, $\phi_{2}(\hat{q})$. Employing the limit,
$\omega\approx m$ on RHS, these equations can be expressed as:
\begin{eqnarray}
 &&\nonumber \left[\frac{M^{2}}{4}-m^{2}-\widehat{q}^{2}\right]\phi_{1}(\hat{q})=\Theta_{P}\overline{V}(\widehat{q})m\phi_{1}(\hat{q})
 +\frac{\Theta_{P}^{2}}{4}\overline{V}^{2}(\widehat{q})\phi_{1}(\hat{q})\\&&
 \left[\frac{M^{2}}{4}-m^{2}-\widehat{q}^{2}\right]\phi_{2}(\hat{q})=\Theta_{P}\overline{V}(\widehat{q})m\phi_{2}(\hat{q})
 +\frac{\Theta_{P}^{2}}{4}\overline{V}^{2}(\widehat{q})\phi_{2}(\hat{q}).
\end{eqnarray}
It is to be mentioned that these two decoupled Eqs. (22) would
resemble h.o. equations, but for $\overline{V}^{2}(\widehat{q})$
term on the RHS of these equations. It will be shown later from
the definition of kernel in Eq.(31), and Eq.(34), that this second
term involving $\overline{V}^{2}(\widehat{q})$ is negligible in
comparison to the first term involving $\overline{V}(\widehat{q})$
on the RHS, and can be dropped, and these equations would then
resemble exact harmonic oscillator equations. However, due to
identical nature of these equations, their solutions are written
as: $\phi_{1}(\hat{q})=\phi_{2}(\hat{q})\approx\phi_{P}(\hat{q})$,
which represent the eigenfunction of the pseudoscalar meson
obtained by solving the full Salpeter equation and we can write
the wave function for $(J^{PC}=0^{-+})$ state as:
\begin{equation}
\psi^{P}(\hat{q}) \approx
[M+\slashed{P}+\frac{\slashed{\hat{q}}\slashed{P}}{m}]\phi_{P}(\hat{q})
\gamma_{5}.
\end{equation}
In view of the arguments made above, in the structure of
$\psi^{P}(\hat{q})$ for pseudoscalar mesons, $M\gamma_{5}$ would
have dominant contribution among all the other Dirac structures  (
and this is more so for heavy $Q\overline{Q}$ mesons)

\item For vector mesons,
the complete decomposition of 4D BS wave function in terms of
various Dirac structures is \cite{smith69}:

\begin{eqnarray}
&&\nonumber\
\Psi^{V}(P,q)=\slashed{\epsilon}\chi_{1}+\slashed{\epsilon}\slashed{P}
\chi_{2} +[q\cdot\epsilon-\slashed{\epsilon}\slashed{
q}]\chi_{3}+[2q.\epsilon\slashed{P}+\slashed{\epsilon}(\slashed{P}\slashed{q}-\slashed{q}\slashed{P})]\chi_{4}\\&&
\
(q.\epsilon)\chi_{5}+(q.\epsilon)\slashed{P}\chi_{6}+(q.\epsilon)\slashed{q}\chi_{7}+
(q.\epsilon)(\slashed{P}\slashed{q}-\slashed{q}\slashed{P})\chi_{8}
\end{eqnarray}
where
$\chi_{\alpha}=\chi_{\alpha}(q^{2},q.P,P^{2});(\alpha=1,2,3,...,8) $
are the Lorentz scalar amplitudes multiplying the various Dirac
structures in the BS wave function, $\Psi(P,q)$. We again mention
that in our framework, by use of a naive power counting rule in
\cite{bhatnagar06,bhatnagar11,bhatnagar14}, we had shown that the
Dirac structures associated with amplitudes $\chi_{1}$ and
$\chi_{2}$ are of leading order. Those with amplitudes
$\chi_3,...,\chi_6$ are sub-leading. We had also shown that, the
structures associated with $\chi_{7}$ and $\chi_{8}$ are more much
more suppressed than even the sub-leading Dirac structures, and
would contribute very little to meson observable (specially heavy
mesons) calculations in comparison to Dirac structures associated
with $\chi_{1}$,...,$\chi_{6}$. Thus we ignore the Dirac structures
associated with $\chi_{7}$ and $\chi_{8}$ (see \cite{bhatnagar14}
for details). And in various calculations
\cite{bhatnagar06,bhatnagar14}, we had shown that among the two
leading Dirac structures associated with amplitudes $\chi_{1}$, and
$\chi_{2}$, the structure associated with $\chi_{1}$ i.e.
$\gamma.\epsilon$ is most dominant. In Ref. \cite{munczek,sauli}, it
was also shown that  $\chi_{1}$ is the dominant amplitude for not
only ground state vector mesons, but also for their higher
excitations.

 We thus write the instantaneous BS wave function
$\psi^{V}(\hat{q})$ with dimensionality $M$ for vector quarkonia,
$(J^{PC}=1^{--})$ up to sub-leading order (i.e. $O(1/M^{1})$ as per
the power counting scheme \cite{bhatnagar14}) as in case of
pseudoscalar mesons, in the center of mass frame as:
\begin{eqnarray}
&&\nonumber\
\psi^{V}(\hat{q})=M\slashed{\epsilon}\chi_{1}+\slashed{\epsilon}\slashed{P}
\chi_{2}+[\slashed{\epsilon}\slashed{\hat{q}}-\hat{q}.\epsilon]\chi_{3}
+[\slashed{P}\slashed{\epsilon}\slashed{\hat{q}}-(\hat{q}.\epsilon)\slashed{P}]\frac{1}{M}\chi_{4}+(\hat{q}.\epsilon)\chi_{5}\\&&
 +(\hat{q}.\epsilon)\slashed{P}\frac{\chi_{6}}{M}
\end{eqnarray}
where $\chi_{1}, ... ,\chi_{6}$ are even functions of $\hat{q}$
and $M$ is the mass of the bound state (of the corresponding
meson). We now derive the mass spectral coupled equations from
Eq.(16). Putting Eq.(25) into the last two equations of Eq.(16),
and we obtain the independent constraints on the components for
the Instantaneous BS wave function:
\begin{eqnarray}
&&\nonumber \chi_{5}=\frac{M\chi_{1}}{m}\\&& \nonumber
\chi_{4}=\frac{-\chi_{2}M}{m},\\&&
 \end{eqnarray}
with $\chi_{6}=\chi_{3}=0$. Applying the constraints in Eq.(26) to
Eq.(26), we can rewrite the relativistic wave function of state
$(J^{PC}=1^{--})$ as:
\begin{equation}
\psi^{V}(\hat{q}) =
[M\slashed{\epsilon}+\hat{q}.\epsilon\frac{M}{m}]\chi_{1}(\hat{q})+[\slashed{\epsilon}\slashed{P}+
\frac{2\slashed{P}\hat{q}.\epsilon}{m}-\frac{\slashed{P}\slashed{\epsilon}\slashed{\hat{q}}}{m}]\chi_{2}(\hat{q}),
\end{equation}
where, we have been able to express the instantaneous wave
function $\psi^{V}(\widehat{q})$ in terms of only the leading
Dirac structures associated with amplitudes $\chi_{1}$ and
$\chi_{2}$. Putting the wave function Eq.(27) into the first two
equations of Eq.(16) and by evaluating trace over the
$\gamma$-matrices on both sides, we can obtain two independent BS
coupled integral equations:
\begin{eqnarray}
 &&\nonumber(M-2\omega)\left[\chi_{1}(\hat{q})
 -\frac{\omega}{m}\chi_{2}(\hat{q})\right]=\Theta_{V}\int\frac{d^{3}\hat{q}'}{(2\pi)^{3}}
 V(\hat{q},\hat{q}')\left[\chi_{1}(\hat{q}')
 -\frac{\omega}{m} \chi_{2}(\hat{q}')\right]\\&&
  \nonumber(M+2\omega)\left[\chi_{1}(\hat{q})
 +\frac{\omega}{m}\chi_{2}(\hat{q})\right]=\Theta_{V}\int\frac{d^{3}\hat{q}'}{(2\pi)^{3}}
 V(\hat{q},\hat{q}')\left[-\chi_{1}(\hat{q}')
 -\frac{\omega}{m} \chi_{2}(\hat{q}')\right]\\&&
\end{eqnarray}
To decouple these equations, we proceed as in the above case of
pseudoscalar mesons. We first add these above equations. Then
subtract the second equation from the first. And for a kernel
expressed as
$V(\widehat{q}-\widehat{q}')=\overline{V}(\widehat{q})\delta^{3}(\widehat{q}-\widehat{q}')$,
we get two algebraic coupled equations in $\chi_{1}$ and
$\chi_{2}$. We Eliminate $\chi_{1}$ from the first equation in
terms of $\chi_{2}$, and plug it in the second equation to get an
equation entirely in terms of $\chi_{2}$. Similarly, we eliminate
$\chi_{2}$ from the second equation in terms of $\chi_{1}$, and
plugging in the first equation to get an equation entirely in
terms of $\chi_{1}$. We thus get two identical decoupled
equations, one in $\chi_{1}(\hat{q})$, and the other in
$\chi_{2}(\hat{q})$, and in the approximation, $\omega\approx m$,
(valid for heavy quarkonia), these equations can be expressed as:
\begin{eqnarray}
 &&\nonumber \left[\frac{M^{2}}{4}-m^{2}-\widehat{q}^{2}\right]\chi_{1}(\hat{q})=\Theta_{V}
 \overline{V}(\widehat{q})m\chi_{1}(\hat{q})+\frac{\Theta_{V}^{2}}{4}\overline{V}^{2}(\widehat{q})\chi_{1}(\hat{q})\\&&
 \left[\frac{M^{2}}{4}-m^{2}-\widehat{q}^{2}\right]\chi_{2}(\hat{q})=\Theta_{V}
 \overline{V}(\widehat{q})m\chi_{2}(\hat{q})+\frac{\Theta_{V}^{2}}{4}\overline{V}^{2}(\widehat{q})\chi_{2}(\hat{q}),
\end{eqnarray}
It is to be mentioned that these two decoupled equations would
again resemble h.o. equations, but for
$\overline{V}^{2}(\widehat{q})$, term on the RHS of these
equations. It will be shown later from the definition of spatial
part V of the kernel in Eq.(31), and Eq.(34), that this term is
negligible in comparison to the first term involving
$\overline{V}$ on the RHS, and can be dropped, and these equations
would then resemble exact harmonic oscillator equations.

From the identical nature of both these equations, we can again
express the solutions as,
$\chi_{1}(\hat{q})=\chi_{2}(\hat{q})\approx \phi_{V}(\hat{q})$, and
we can write the wave function for $(J^{PC}=1^{--})$ as:
\begin{equation}
\psi^{V}(\hat{q})\approx
[M\slashed{\epsilon}+\hat{q}.\epsilon\frac{M}{m}+\slashed{\epsilon}\slashed{P}+
\frac{2\slashed{P}\hat{q}.\epsilon}{m}-\frac{\slashed{P}\slashed{\epsilon
}\slashed{\hat{q}}}{m}]\phi_{V}(\hat{q}).
\end{equation}
\end{itemize}

In the structure of $\psi^{V}(\hat{q})$ for vector meson, as argued
above, $M\slashed{\epsilon}$ is the most dominant Dirac structure,
among all the other Dirac structures above (and this is more so for
heavy $Q\overline{Q}$ mesons). We wish to mention that the framework
is quite general up to this point. We now introduce the BS kernel.
\bigskip

Now, as regards the BS kernel $K(q,q')$
\cite{mitra01,bhatnagar06,bhatnagar11,bhatnagar14} is concerned,
it is taken to be one-gluon-exchange like as regards the spin
dependence $(\gamma_{\mu}\bigotimes\gamma^{\mu})$, and color
dependence
$(\frac{1}{2}\vec{\lambda}_{1}.\frac{1}{2}\vec{\lambda}_{2})$, and
has a scalar part $V$:

\begin{eqnarray}
&&\nonumber K(q,q') =
(\frac{1}{2}\vec{\lambda}_{1}.\frac{1}{2}\vec{\lambda}_{2})(\gamma_{\mu}\otimes\gamma^{\mu})V(q-q')\\&&
\nonumber\ V(\hat{q},\hat{q}')=
\frac{3}{4}\omega^{2}_{q\bar{q}}\int
d^{3}\vec{r}[r^{2}(1+4\hat{m}_{1}\hat{m}_{2}A_{0}M_{>}^{2}r^{2})^{-\frac{1}{2}}-\frac{C_{0}}{\omega_{0}^{2}}]
e^{i(\hat{q}-\hat{q}').\vec{r}}\\&&
\nonumber\omega^{2}_{q\bar{q}}=4M_{>}\hat{m}_{1}\hat{m}_{2}\omega^{2}_{0}\alpha_{s}(M_{>}
^{2})\\&&
\nonumber\alpha_{s}(M_{>}^{2})=\frac{12\pi}{33-2n_{f}}[log(\frac{M_{>}^{2}}{\wedge^{2}})]^{-1}\\&&
\nonumber
\hat{m}_{1,2}=\frac{1}{2}[1\pm\frac{(m^{2}_{1}-m^{2}_{2})}{M^{2}}]\\&&
\nonumber
\kappa=(1+4\hat{m}_{1}\hat{m}_{2}A_{0}M^{2}r^{2})^{-\frac{1}{2}}\\&&
\ M_{>}= Max(M, m_{1}+m_{2}).
\end{eqnarray}
The scalar part is purely confining (as in Ref.
\cite{koll,babutsidze,hluf15}, and the Martin potential
\cite{olsson} employed for heavy mesons). Here the proportionality
of $\omega_{q\overline{q}}^{2}$ on $\alpha_{S}(Q^{2})$ is needed
to provide a more direct QCD motivation \cite{mitra01} to
confinement, and $\omega_{0}^{2}$ is postulated as a spring
constant which is common to all flavors. Here in the expression
for $V(\widehat{q},\widehat{q}')$, the constant term
$C_{0}/\omega_{0}^{2}$ is designed to take account of the correct
zero point energies, while the $A_{0}$ term $(A_{0} << 1)$
simulates an effect of an almost linear confinement for heavy
quark sectors (large $m_{1}, m_{2}$), while retaining the harmonic
form for light quark sectors (small $m_{1}, m_{2}$), as is
believed to be true for QCD (see
\cite{mitra01,bhatnagar06,bhatnagar14}). Hence the term
$(1+4\hat{m}_{1}\hat{m}_{2}A_{0}M^{2}r^{2})^{-\frac{1}{2}}$ in the
above expression is responsible for effecting a smooth transition
from harmonic ($q\overline{q}$) to linear ($Q\overline{Q}$)
confinement.
\bigskip

We now try to work on the spatial part $V(\widehat{q},\widehat{q}')$
of the confining potential $K(\widehat{q},\widehat{q}')$. If we take
the parameter $A_{0}=0$ (which corresponds to case of light mesons
($q\overline{q}$), since due to $A_{0}<<0$, the square root factor
in the denominator,
$\kappa=(1+4\hat{m}_{1}\hat{m}_{2}A_{0}M^{2}r^{2})^{-\frac{1}{2}}=1)$,
and $V$ would look like:

\begin{equation}
V'(\hat{q},\hat{q}')= \frac{3}{4}\omega^{2}_{q\bar{q}}\int
d^{3}\vec{r}[r^{2}-\frac{C_{0}}{\omega_{0}^{2}}]
e^{i(\hat{q}-\hat{q}').\vec{r}}.
\end{equation}

(where $V'$ is $V$ without the factor $\kappa$ in denominator).
Making use of the fact that
$-\overrightarrow{\nabla}_{\widehat{q}}^{2}$ is the fourier
transform of $r^{2}$ in momentum space, and the colour factor for
bound $q\overline{q}$ system,
$(\frac{1}{2}\vec{\lambda}_{1}.\frac{1}{2}\vec{\lambda}_{2})=-\frac{4}{3}$,
we can write
\begin{equation}
\
V'(\widehat{q},\widehat{q}')=\omega_{q\overline{q}}^{2}(2\pi)^{3}[\overrightarrow{\nabla}_{\widehat{q}}^{2}+\frac{C_{0}}{\omega_{0}^{2}}]\delta^{3}(\widehat{q}-\widehat{q}').
\end{equation}

which can be written as,
$V'(\widehat{q},\widehat{q}')=\overline{V'}(\widehat{q})\delta^{3}(\widehat{q}-\widehat{q}')$.
Putting this spatial part in place of the spatial part of the
kernel $V$ in Eq.(22) and Eq.(29) (with the second term on the RHS
of each of these equations dropped for reasons explained next),
these equations would resemble harmonic oscillator equations, and
we can have analytical solutions for them. But for parameter
$A_{0}\neq 0$ (i.e if we reintroduce the factor $\kappa$), we can
write the complete potential $V(\widehat{q},\widehat{q}')$ as:
\begin{eqnarray}
&&\nonumber\
V(\hat{q},\hat{q}')=\overline{V}(\hat{q})\delta^{3}(\hat{q}-\hat{q}'),\\&&
\nonumber
\overline{V}(\hat{q})=\omega^{2}_{q\bar{q}}[\kappa\overrightarrow{\nabla}^{2}_{\hat{q}}
+\frac{C_{0}} {\omega^{2}_{0}}](2\pi)^{3},\\&&
\kappa=(1-A_{0}M^{2}\overrightarrow{\nabla}^{2}_{\hat{q}})^{-1/2}
\end{eqnarray}
\bigskip
As mentioned above, we have dropped the terms with
$\overline{V}^{2}$, in comparison to the terms involving
$\overline{V}$ on the RHS of Eqs. (22) and (29), as the
coefficients, $\Omega_{P,V}'=
\frac{\Theta_{P,V}^2}{4}\omega^{4}_{q\bar{q}}$ associated with the
former have a very small contribution $(\leq 0.638\%)$ in
comparison to the coefficient
$\Omega_{P,V}=m\Theta_{P,V}\omega^{2}_{q\bar{q}}$ associated with
the latter for both pseudoscalar and vector mesons, due to
$\omega^{4}_{q\bar{q}}<< \omega^{2}_{q\bar{q}}$ for $\eta_{c}$,
$\eta_{b}$, $J/\psi$ and $\Upsilon$. The numerical values of these
coefficients, and their percentage ratio for both pseudoscalar
($\eta_{c}$, $\eta_{b}$), and vector ($J/\psi$ and $\Upsilon$)
mesons are given in Table 2 below, which justifies these terms
being dropped.

\begin{table}[h]
  \begin{center}
\begin{tabular}{lllllll}
  \hline
     &$\Omega_{P}$&$\Omega_{P}'$&$\frac{\Omega_{P}'}{\Omega_{P}}\%$&$\Omega_{V}$&$\Omega_{V}'$&
     $\frac{\Omega_{V}'}{\Omega_{V}}\%$\\\hline
    $\eta_{c}$&0.0558&0.000356&0.638\%& & & \\
    $\eta_{b}$&0.4564&0.00202&0.4426\%& & & \\
$J/\psi$& & & &0.0246&0.00006889&0.280\%\\ $\Upsilon$&
&&&0.2293&0.0005115&0.2230\%\\\hline\hline
\end{tabular}
\end{center}
\caption{Numerical values of coefficients,
$\Omega_{P}=m\Theta_{P}\omega^{2}_{q\bar{q}}$, and $\Omega_{P}'=
\frac{\Theta_{P}^2}{4}\omega^{4}_{q\bar{q}}$ associated with the
terms involving $\overline{V}$ and $\overline{V}^{2}$
respectively, for pseudoscalar mesons $\eta_{c}$ and $\eta_{b}$ in
RHS of Eqs.(22), and their percentage ratio, along with the
corresponding values, and percentage ratio of
$\Omega_{V}=m\Theta_{V}\omega^{2}_{q\bar{q}}$, and $\Omega_{V}'=
\frac{\Theta_{V}^2}{4}\omega^{4}_{q\bar{q}}$ for vector mesons,
$J/\psi$ and $\Upsilon$ on RHS of Eqs.(29). The input parameters
of our model are: $C_{0}=0.21$, $\omega_{0}=.15GeV.$, QCD length
scale $\Lambda=0.200GeV.$, $A_{0}=0.01$, and the input quark
masses, $m_{c}=1.49GeV.$, and $m_{b}=5.070GeV.$}
    \end{table}
\bigskip
To derive the mass spectrum, we put the spatial part,
$\overline{V}$ in Eq.(34) into the equations Eq.(22), and Eq.(29)
for pseudoscalar and vector mesons respectively. But the form of
$V$ in Eq.(34) suggests that these equations have to be solved
numerically. However, to solve these equations with $A_{0}\neq 0$,
we follow an analytical procedure on lines of
Ref.\cite{mitra99,mitra89}, where we treat $\kappa$ as a
"correction" factor due to small value of parameter, $A_{0}$
($A_{0}<< 1$), while we work in an approximate harmonic oscillator
basis, due to its transparency in bringing out the dependence of
mass spectral equations on the total quantum number $N$. (In this
connection, we wish to mention that recently, harmonic oscillator
basis has also been widely employed to study heavy quarkonia using
a Light-front quark model \cite{peng12}.) The latter is achieved
through the effective replacement,
$\kappa=(1-A_{0}M^{2}\overrightarrow{\nabla}_{\hat{q}}^{2})^{-\frac{1}{2}}\Rightarrow
(1+2A_{0}(N+\frac{3}{2}))^{-\frac{1}{2}}$, (which is quite valid
for heavy $c\overline{c}$ and $b\overline{b}$ systems), in Eq.(34)
on lines of \cite{mitra99,mitra89}. With this, we can reduce
Eq.(22) and Eq.(29) to equations of a simple quantum mechanical
3D- harmonic oscillator with coefficients depending on the hadron
mass $M$ and total quantum number $N$. The wave function satisfies
the 3D BSE for equal mass heavy pseudoscalar and vector mesons
respectively as given below:
\bigskip
\begin{equation}
(\frac{M^{2}}{4}-m^{2}-\widehat{q}^{2})\phi_{P}(\hat{q})=\Theta_{P}m\omega^{2}_{q\bar{q}}
 [\frac{\overrightarrow{\nabla}^{2}_{\hat{q}}}
{\sqrt{1+2A_{0}(N+\frac{3}{2})}}+\frac{C_{0}}
{\omega^{2}_{0}}]\phi_{P}(\hat{q})
 \end{equation}
 and

\begin{equation}
 (\frac{M^{2}}{4}-m^{2}-\widehat{q}^{2})\phi_{V}(\hat{q})=\Theta_{V}
 m\omega^{2}_{q\bar{q}}[\frac{\overrightarrow{\nabla}^{2}_{\hat{q}}}
{\sqrt{1+2A_{0}(N+\frac{3}{2})}}+\frac{C_{0}}
{\omega^{2}_{0}}]\phi_{V}(\hat{q}),
\end{equation}
\bigskip

where, with use of dominant Dirac structures
\cite{bhatnagar11,bhatnagar14}, we can to a good approximation
express, $\Theta_{P}=-4$ for pseudoscalar states, and
$\Theta_{V}=-2$ for vector states, and write Eqs.(35) and (36) in
the same expression as:
\begin{equation}
E_{P,V}\phi_{P,V}(\hat{q})=(-\beta_{P,V}^{4}\overrightarrow{\nabla}^{2}_{\hat{q}}+\hat{q}^{2})\phi_{P,V}(\hat{q}),
\end{equation}
where,
$\overrightarrow{\nabla}_{\widehat{q}}^{2}=\frac{\partial^{2}}{\partial\widehat{q}^{2}}+\frac{2}{\widehat{q}}\frac{\partial}{\partial\widehat{q}}-\frac{l(l+1)}{\widehat{q}^{2}}$,
and $l=0,1,2,...$ correspond to $S,P,D,...$ states respectively.
$E_{P,V}=\frac{M^{2}}{4}-m^{2}+\frac{\beta_{P,V}^{4}C_{0}}{\omega^{2}_{0}}\sqrt{1+2A_{0}(N+\frac{3}{2})}$,
and $\phi_{P,V}(\hat{q})$ are the eigen functions of Eq.(37) for
equal mass heavy pseudoscalar, and vector quarkonia. The inverse
range parameters for pseudoscalar and vector meson respectively
are,
$\beta_{P}=(4\frac{m\omega^{2}_{q\bar{q}}}{\sqrt{1+2A_{0}(N+\frac{3}{2})}})^{\frac{1}{4}}$,
and
$\beta_{V}=(2\frac{m\omega^{2}_{q\bar{q}}}{\sqrt{1+2A_{0}(N+\frac{3}{2})}})^{\frac{1}{4}}$
and are dependent on the input kernel and contains the dynamical
information, and which only differ from each other due to
spin-spin interactions. We can express Eq.(37) for the
$l=0,2,...,$ states studied here as:
\begin{equation}
\phi_{P,V}''(\widehat{q})
+\frac{2}{\widehat{q}}\phi_{P,V}'(\widehat{q})
+\frac{1}{\beta_{P,V}^{4}}[E_{P,V}-\frac{l(l+1)\beta_{P,V}^{4}}{\widehat{q}^{2}}-\widehat{q}^{2}]\phi_{P,V}(\widehat{q})=0.
\end{equation}

Assuming the form of the solutions of the above equation as
$\phi_{P,V}(\widehat{q})=h(\widehat{q})\widehat{q}^{l}e^{-\frac{\hat{q}^{2}}{2\beta^{2}}}$,
where $\beta^{2}=\beta_{P,V}^{2}$, we obtain the equation,
\begin{equation}
h''(\widehat{q})+(\frac{2l}{\widehat{q}}+\frac{2}{\widehat{q}}-\frac{2\widehat{q}}{\beta^{2}})h'(\widehat{q})
+[\frac{E}{\beta^{4}}-\frac{2l}{\beta^{2}}-\frac{3}{\beta^{2}}]h(\widehat{q})=0.
\end{equation}
Making use of the power series method to solve this equation, we
obtain the energy eigenvalues as:

\begin{equation}
(E_{P,V})_{N}=2\beta^{2}(N+\frac{3}{2}); N=2n+l; n=0,1,2,...,
\end{equation}
with the normalized forms of energy eigen functions for $l=0 (S)$,
and for $l=2 (D)$ states derived as:
\begin{eqnarray}
&&\nonumber
\phi_{P(V)}(1S,\hat{q})=\frac{1}{\pi^{3/4}\beta^{3/2}}e^{-\frac{\hat{q}^{2}}{2\beta^{2}}},\\&&
\nonumber \phi_{P(V)}(2S,\hat{q})=
(\frac{3}{2})^{1/2}\frac{1}{\pi^{3/4}\beta^{3/2}}(1-\frac{2\hat{q}^{2}}{3\beta^{2}})e^{-\frac{\hat{q}^{2}}{2\beta^{2}}},\\&&
\nonumber
\phi_{V}(1D,\hat{q})=(\frac{4}{15})^{1/2}\frac{1}{\pi^{3/4}\beta^{7/2}}\widehat{q}^{2}e^{-\frac{\hat{q}^{2}}{2\beta^{2}}}\\&&
\nonumber
\phi_{P(V)}(3S,\hat{q})=(\frac{15}{8})^{1/2}\frac{1}{\pi^{3/4}\beta^{3/2}}
(1-\frac{20\hat{q}^{2}}{15\beta^{2}}+\frac{4\hat{q}^{4}}{15\beta^{4}})e^{-\frac{\hat{q}^{2}}{2\beta^{2}}},\\&&
\nonumber
\phi_{V}(2D,\hat{q})=(\frac{14}{15})^{1/2}\frac{1}{\pi^{3/4}\beta^{7/2}}
(1-\frac{2\widehat{q}^{2}}{7\beta^{2}}) \widehat{q}^{2}
e^{-\frac{\hat{q}^{2}}{2\beta^{2}}},\\&& \nonumber
\phi_{P(V)}(4S,\hat{q})=(\frac{35}{16})^{1/2}\frac{1}{\pi^{3/4}\beta^{3/2}}
(1-\frac{210\hat{q}^{2}}{105\beta^{2}}+\frac{84\hat{q}^{4}}{105\beta^{4}}-\frac{8\widehat{q}^{6}}{105\beta^{6}})
e^{-\frac{\hat{q}^{2}}{2\beta^{2}}}\\&&
\phi_{V}(3D,\hat{q})=(\frac{21}{10})^{1/2}\frac{1}{\pi^{3/4}\beta^{7/2}}
(1-\frac{36\hat{q}^{2}}{63\beta^{2}}+\frac{4\hat{q}^{4}}{63\beta^{4}})\hat{q}^{2}e^{-\frac{\hat{q}^{2}}{2\beta^{2}}}.
\end{eqnarray}

We will use them for a description of the equal mass heavy
pseudoscalar and vector mesons. In Fig.1, and Fig.2, we now give
the plots of these normalized wave functions Vs. $\widehat{q}$ (in
Gev.) for different states of pseudoscalar $c\overline{c}$, and
$b\overline{b}$ quarkonia. And in Fig.3, and Fig.4, we give the
plots of these wave functions Vs. $\widehat{q}$ (in Gev.) for
different states of vector $c\overline{c}$, and $b\overline{b}$
quarkonia. It can be seen from these plots that the wave functions
corresponding to $nS$, and $nD$ states have $n-1$ nodes.

\begin{figure}[h]
\centering
\includegraphics[width=14cm]{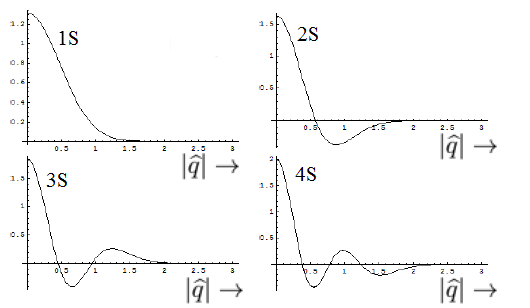}
\caption{Plots of wave functions for states $(1S,...,4S)$ Vs
$\widehat{q}$ (in Gev.) for pseudoscalar $c\overline{c}$ states.}
\end{figure}

\begin{figure}[h]
\centering
\includegraphics[width=14cm]{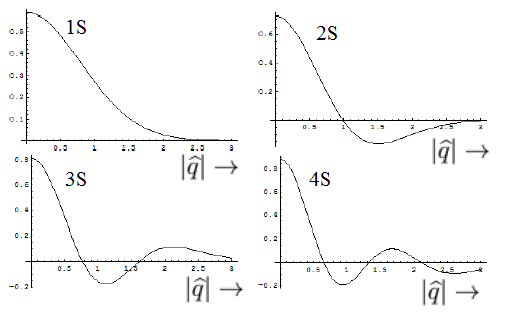}
\caption{Plots of wave functions for states $(1S,...,4S)$ Vs
$\widehat{q}$ (in Gev.) for pseudoscalar $b\overline{b}$ states.}
\end{figure}

\begin{figure}[h]
\centering
\includegraphics[width=16cm]{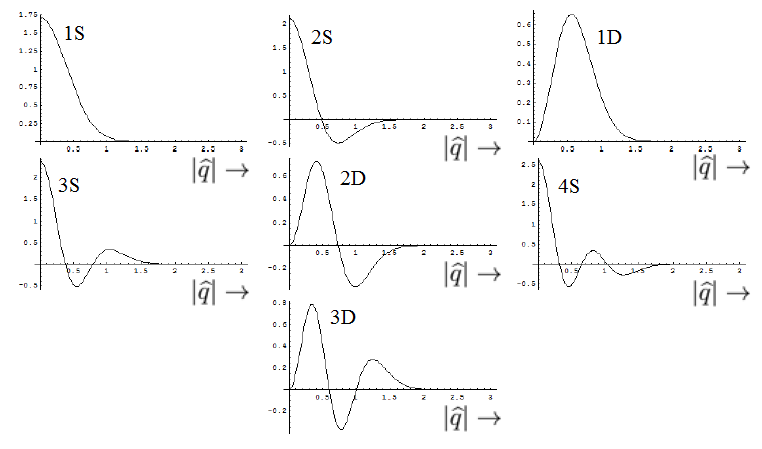}
\caption{Plots of wave functions for states $(1S,...,3D)$ Vs
$\widehat{q}$ (in Gev.) for vector $c\overline{c}$ states.}
\end{figure}

\begin{figure}[h]
\centering
\includegraphics[width=16cm]{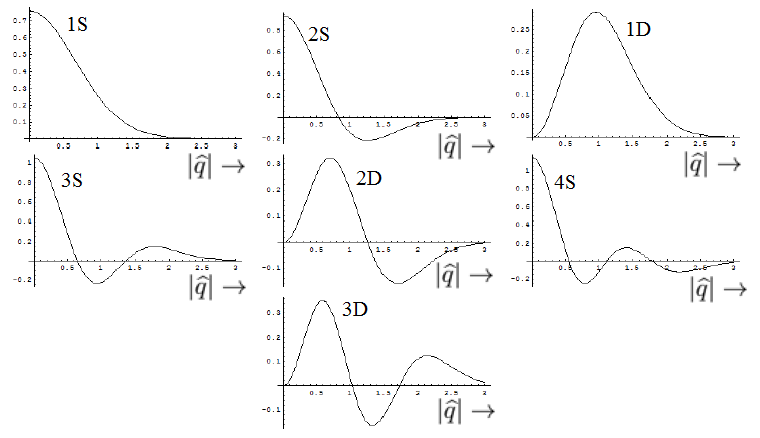}
\caption{Plots of wave functions for states $(1S,...,3D)$ Vs
$\widehat{q}$ (in Gev.) for vector $b\overline{b}$ states.}
\end{figure}

The mass spectrum of ground (1S) and excited states for equal mass
heavy pseudoscalar ($0^{-+}$) and vector ($1^{--}$) mesons
respectively is written as:
\begin{equation}
\frac{1}{2\beta_{P}^{2}}(\frac{M^{2}}{4}-m^{2}+\frac{C_{0}\beta_{P}^{4}}{\omega^{2}_{0}}
 \sqrt{1+2A_{0}(N+\frac{3}{2})})=N+\frac{3}{2}; N=2n+l; n=0,1,2,...,
\end{equation}
and
\begin{equation}
\frac{1}{2\beta_{V}^{2}}(\frac{M^{2}}{4}-m^{2}+\frac{C_{0}\beta_{V}^{4}}{\omega^{2}_{0}}
 \sqrt{1+2A_{0}(N+\frac{3}{2})})=N+\frac{3}{2}; N=2n+l; n=0,1,2....
\end{equation}

\begin{table}[htbp]
\begin{center}
\begin{tabular}{lllllll}
  \hline
    &BSE - CIA &Expt.\cite{olive14}&Pot.
    Model\cite{bhagyesh11}&
   QCD sum rule\cite{veli12}&Lattice QCD\cite{burch09}&\cite{ebert13} \\\hline
    $M_{\eta_{c}(1S)}$& 2.9509 & 2.983$\pm$0.0007&2.980 & 3.11$\pm$0.52 &3.292&2.981 \\
    $M_{\eta_{c}(2S)}$& 3.7352 & 3.639$\pm$0.0013  &3.600& &4.240&3.635\\
    $M_{\eta_{c}(3S)}$& 4.4486 &   &4.060& &&3.989\\
    $M_{\eta_{c}(4S)}$& 5.1048&    &4.4554& &&4.401\\
    $M_{\eta_{b}(1S)}$& 9.0005 & 9.398 $\pm$0.0032
    &9.390 &9.66$\pm$ 1.65&7.377&9.398  \\
    $M_{\eta_{b}(2S)}$& 9.7215 & 9.999$\pm$0.0028   &9.947 & &8.202&9.990   \\
    $M_{\eta_{b}(3S)}$&10.4201 &   & 10.291& &&10.329\\
    $M_{\eta_{b}(4S)}$&11.0968&     &      &  &&10.573\\\hline
    \hline
     \end{tabular}
   \end{center}
   \caption{Masses of ground and radially excited
states of $\eta_c$ and $\eta_b$ (in GeV.) in present calculation
(BSE-CIA) along with experimental data, and their masses in other
models.}
\end{table}
\begin{table}[htbp]
  \begin{center}
\begin{tabular}{lllllll}
  \hline
    &BSE - CIA &Expt.\cite{olive14}&Rel. Pot. Model\cite{ebert13}&Pot. Model\cite{bhagyesh11}
    &BSE\cite{wang06}&Lattice QCD\cite{kawanai15}\\\hline
    $M_{J/\psi(1S)}$&3.0974& 3.0969$\pm$ 0.000011&3.096       & 3.0969 &3.0969&3.099 \\
    $M_{\psi(2S)}$&3.6676& 3.6861$\pm$ 0.00034  &3.685& 3.6890&3.686 &3.653\\
    $M_{\psi(1D)}$&3.6676&3.773$\pm$ 0.00033&3.783  & &3.759&\\
    $M_{\psi(3S)}$&4.1945&4.03$\pm$ 0.001&4.039       & 4.1407&4.065 &4.099\\
    $M_{\psi(2D)}$&4.1945&4.191$\pm$0.005&4.150 &&4.108 &\\
    $M_{\psi(4S)}$& 4.6856&4.421$\pm$0.004 &4.427&4.5320&4.344& \\
    $M_{\psi(3D)}$& 4.6856&&4.507 && 4.371  & \\
    $M_{\psi(5S)}$& 5.1463&& 4.837   &4.8841 & 4.567 &    \\
    $M_{\psi(4D)}$& 5.1463&&4.857    & &  &    \\
    $M_{\Upsilon(1S)}$&9.6719&9.4603$\pm$ 0.00026 &9.460&9.4603 &9.460&\\
    $M_{\Upsilon(2S)}$&10.1926&10.0233$\pm$0.00031&10.023&9.9814 &10.029 &\\
    $M_{\Upsilon(1D)}$&10.1926& &10.154&& 10.139&\\
    $M_{\Upsilon(3S)}$&10.6979&10.3552$\pm$0.00005&10.355&10.3195 &10.379 &\\
    $M_{\Upsilon(2D)}$&10.6979&&10.435&& 10.438&\\
    $M_{\Upsilon(4S)}$&11.1887&10.5794$\pm$0.0012&10.586  &10.5995 &10.648& \\
    $M_{\Upsilon(3D)}$&11.1887&&10.704&& 10.690& \\
    $M_{\Upsilon(5S)}$&11.6657&10.865$\pm$0.008&10.869     &10.8465&10.868&\\
    $M_{\Upsilon(4D)}$&11.6657&&10.949     && &\\
    $M_{\Upsilon(6S)}$&12.1296&11.019$\pm$0.008&11.088    &11.0713&    &\\
    $M_{\Upsilon(5D)}$&12.1296&&    && &\\
     \hline
    \hline
     \end{tabular}
   \end{center}
   \caption{Masses of ground, radially and orbitally excited states of heavy vector quarkonium,
        $J/\psi$ and $\Upsilon$ in BSE-CIA along with their masses in other models and experimental
          data (all units are in GeV).}
     \end{table}
The input parameters of our model are: $C_{0}=0.21$,
$\omega_{0}=.15GeV.$, QCD length scale $\Lambda=0.200GeV.$,
$A_{0}=0.01$, and the input quark masses, $m_{c}=1.49GeV.$, and
$m_{b}=5.070GeV.$ The results of mass spectral predictions of
heavy equal mass pseudoscalar and vector mesons for both ground
and
 excited states with the above set of parameters is given in
table 3 and 4. We now present the calculation of the leptonic decay
constants $f_P$ and $f_V$ for these equal mass heavy pseudoscalar
and vector mesons, $\eta_{c}$, $\eta_{b}$, $J/\psi$ and $\Upsilon$.

\section{ Leptonic Decays of equal mass Heavy pseudoscalar and vector quarkonia}
We now do the calculation of decay constants of equal mass
pseudoscalar and vector mesons such as, $\eta_{c}$, $\eta_{b}$,
$J/\psi$ and $\Upsilon$, which are defined as
\cite{hluf15,dudek06},
\begin{eqnarray}
&&\nonumber if_{P}P_{\mu} \equiv<0 |\bar{Q}i\gamma_{\mu}\gamma_{5}Q|
P>.\\&&
 \ f_{V}M\epsilon_{\mu}(P) \equiv<0
|\bar{Q}\gamma_{\mu}Q| V(P)>
\end{eqnarray}
The decay constants $f_{P}$ and $f_{V}$ thus can be evaluated
through the loop diagram which gives the coupling of  two-quark
loop to the axial vector current and vector current respectively,
and can be expressed as a quark-loop integral (for some of our
recent works on leptonic decays in the framework of BSE, see,
\cite{bhatnagar06,bhatnagar11,bhatnagar14,hluf15}):
\begin{eqnarray}
&&\nonumber \ f_{P}P_{\mu}=\sqrt{3}\int \frac{d^{4}q}{(2\pi)^{4}}
Tr[\Psi^{P}(P,q)i\gamma_{\mu}\gamma_{5}],\\&&
 \ f_{V}M\epsilon_{\mu} =\sqrt{3}\int \frac{d^{4}q}{(2\pi)^{4}}
Tr[\Psi^{V}(P,q)i\gamma_{\mu}],
\end{eqnarray}
where $\epsilon_{\mu}$ is polarization vector of vector meson
satisfying $\epsilon.P=0$. These equations can be reduced to 3D
forms by defining the 3D wave function, $\psi(\hat{q}) =
-\int\frac{Md\sigma}{2\pi i}\Psi(P,q)$. Thus we can write Eq.(45)
as:
\begin{eqnarray}
&&\nonumber f_{P}P_{\mu} = \sqrt{3} \int
\frac{d^{3}\hat{q}}{(2\pi)^{3}}
Tr[\psi^{P}(\hat{q})\gamma_{\mu}\gamma_{5}]\\&& \
f_{V}M\epsilon_{\mu} =\sqrt{3}\int \frac{d^{3}\hat{q}}{(2\pi)^{3}}
Tr[\psi^{V}(\hat{q})\gamma_{\mu}]
\end{eqnarray}
where (following Eq.(23) and Eq.(30)), the complete 3D
Bethe-Salpeter wave function of state $(0^{-+})$ and $(1^{--})$ is
rewritten as:
\begin{eqnarray}
&&\nonumber \psi^{P}(\hat{q}) =N_{P}\phi_{P}(\widehat{q})
[M+\slashed{P}+\frac{\slashed{\hat{q}}\slashed{P}}{m}]
\gamma_{5},\\&&
\psi^{V}(\hat{q})=N_{V}\phi_{V}(\hat{q})[M\slashed{\epsilon}+\hat{q}.\epsilon\frac{M}{m}+\slashed{\epsilon}\slashed{P}+
\frac{\slashed{P}\hat{q}.\epsilon}{m}-\frac{\slashed{P}\slashed{\epsilon
}\slashed{\hat{q}}}{m}],
\end{eqnarray}

where $N_{P}$ and $N_{V}$ are the standard  BS normalizers which
enter into the BS wave function, and $m$ and $M$ are the masses of
quarks and the corresponding meson respectively.

The 4D BS normalizer $N_P$ and $N_{V}$ are evaluated from the
current conservation condition:
\begin{equation}
2iP_\mu=\int \frac{d^{4}q}{(2\pi)^{4}}
\mbox{Tr}\left\{\overline{\Psi}(P,q)\left[\frac{\partial}{\partial
P_\mu}S_{F}^{-1}(p_1)\right]\Psi(P,q)S_{F}^{-1}(-p_2)\right\} +
(1\rightleftharpoons2).
\end{equation}
Carrying out derivatives of inverse quark propagators of
constituent quarks with respect to total hadron momentum $P_\mu$,
evaluating trace over products of gamma matrices, following usual
steps, we then express the above equation in terms of the
integration variables $\hat{q}$ and $\sigma$. Noting that the 4D
volume element $d^4q =d^3\widehat{q}Md\sigma$, we then perform the
contour integration in the complex $\sigma$- plane by making use
of the corresponding pole positions. Then integration over the
variable $\hat{q}$ is finally performed to extract out the
numerical results for $N_P$ and $N_{V}$ for different equal mass
pseudoscalar and vector mesons. The above equation due to the
orthogonality condition, $P.\widehat{q}=0$ and  $P.\epsilon=0$
reduces to a simple forms given below for equal mass pseudoscalar
and vector mesons:
\begin{eqnarray}
 N_{P}^{-2} =\frac{16M}{m}\int\frac{d^{3}\hat{q}}{(2\pi)^{3}}
\frac{\hat{q}^{2}}{\omega}\phi_{P}^{2}(\hat{q})
\end{eqnarray} and
\begin{eqnarray}
N_{V}^{-2} =16Mm\int \frac{d^{3}\hat{q}}{(2\pi)^{3}}
\frac{\hat{q}^{2}}{\omega^{3}}\phi_{V}^{2}(\hat{q}).
\end{eqnarray}

Here, in the above equations for decay constants and the BS
normalizers, the wave functions, $\phi_{P}$ and $\phi_{V}$
represent the eigenfunctions of the pseudoscalar and vector mesons
obtained by solving the full Salpeter equation. Their algebraic
expressions are taken from Eq. (41), and are then employed to
calculate $f_{P}$ and $f_{V}$ as well as the BS normalizers
$N_{P}$, and $N_{V}$ for ground as well as the excited states of
$\eta_{c}, \eta_{b}, J/\psi$, and $\Upsilon$. The numerical values
of BS normalizers of all these states are given in Table 5 below.

\begin{table}[htbp]
\begin{center}
\begin{tabular}{lllll}
  \hline
   &$N_{\eta_{c}}$ &$N_{\eta_{b}}$&$N_{J/\psi}$&$N_{\Upsilon}$\\\hline
    $1S$&5.2239& 5.7441& 5.9102& 6.4666\\
    $2S$&3.1632& 3.6679 & 3.6788& 4.1743  \\
    $1D$& &  & 3.6233  &4.1534 \\
    $3S$&2.3845&2.8657 & 2.8317 & 3.2904 \\
    $2D$&      &       &2.8079     &3.2805  \\
    $4S$&2.1531&2.4103  &2.3591      &2.7881\\
    $3D$&   &     & 2.3462&2.7058  \\
   \hline
   \end{tabular}
   \end{center}
   \caption{Numerical values of BS normalizers for ground
state and excited states of $\eta_{c}$, $\eta_{b}$, $J/\psi$ and
$\Upsilon$ (in GeV units) in present calculation.}
     \end{table}

The decay constant from Eq.(46) becomes:
\begin{eqnarray}
&&\nonumber \ f_{P}P_{\mu} = \sqrt{3} N_{P}\int
\frac{d^{3}\hat{q}}{(2\pi)^{3}}\phi_{P}(\hat{q}) Tr[
(M+\slashed{P}+\frac{\slashed{\hat{q}}\slashed{P}}{m})\gamma_{\mu}]\\&&
\nonumber f_{V}M\epsilon_{\mu}=\sqrt{3} N_{V}\int
\frac{d^{3}\hat{q}}{(2\pi)^{3}}\phi_{V}(\hat{q}) Tr[
(M\slashed{\epsilon}+\hat{q}.\epsilon\frac{M}{m}+\slashed{\epsilon}\slashed{P}+
\frac{\slashed{P}\hat{q}.\epsilon}{m}-\frac{\slashed{P}\slashed{\epsilon
}\slashed{\hat{q}}}{m})\gamma_{\mu}]\\&&
\end{eqnarray}
Evaluating  trace  over $\gamma$-matrices, and carrying out
integration over $d^{3}\hat{q}$, we obtain:
\begin{eqnarray}
&&\nonumber \ f_{P}= 4\sqrt{3}N_{P}\int
\frac{d^{3}\hat{q}}{(2\pi)^3}\phi_{P}(\hat{q})\\&&
\
f_{V}=4\sqrt{3}N_{V}\int
\frac{d^{3}\hat{q}}{(2\pi)^{3}}\phi_{V}(\hat{q}).
\end{eqnarray}

\begin{table}[htbp]
\begin{center}
\begin{tabular}{llllll}
  \hline
   &BSE - CIA &Expt.\cite{cleo01}&Lattice QCD \cite{mcnielle12}&Pot. Model\cite{bhagyesh11}&QCD sum rule\cite{veli12}\\\hline
    $f_{\eta_{c}(1S)}$&0.4044& 0.335$\pm$0.075& 0.3928& 0.471& 0.260$\pm$0.075\\
    $f_{\eta_{c}(2S)}$&0.3308&  &      &0.374 & \\
    $f_{\eta_{c}(3S)}$&0.2908&  &      &0.332 & \\
    $f_{\eta_{b}(1S)}$&1.0168&  & 0.667 & 0.834&0.251$\pm$0.072 \\
    $f_{\eta_{b}(2S)}$&0.8066&  &      &0.567 &   \\
    $f_{\eta_{b}(3S)}$&0.7134&  &      &0.508  & \\ \hline
   \hline
   \end{tabular}
   \end{center}
   \caption{Leptonic decay constants, $f_P$ of ground
state (1S) and excited state (2S) and (3S) of $\eta_{c}$ and
$\eta_{b}$ (in GeV.) in present calculation (BSE-CIA) along with
experimental data, and their masses in other models.}
     \end{table}
\begin{table}[htbp]
  \begin{center}
\begin{tabular}{lllllll}
  \hline
     &BSE - CIA &Expt.\cite{olive14}&Pot. Model\cite{bhagyesh11}&BSE \cite{wang06}&LatticeQCD\cite{dudek06}
     &Light front model\cite{peng12}\\\hline
    $f_{J/\psi(1S)}$& 0.3745&0.411$\pm$.007&0.317 &0.459$\pm$.028&0.399$\pm$ 0.004& \\
    $f_{\psi(2S)}$& 0.2953&0.279$\pm$.008&0.253&0.364$\pm$.024&0.143$\pm$ 0.081&0.2474\\
    $f_{\psi(1D)}$&0.2897&0.210$\pm$0.00024&&0.243$\pm$.017&&\\
    $f_{\psi(3S)}$& 0.2610&0.174$\pm$.018&0.226&0.319$\pm$.022& &\\
    $f_{\psi(2D)}$& 0.2348&0.1424$\pm$0.0033&&0.157$\pm$.011& &\\
    $f_{\psi(4S)}$& 0.2399&0.1608$\pm$0.0016&&0.288$\pm$.018& &\\
    $f_{\psi(3D)}$& 0.2570&0.1424$\pm$0.0033&&0.157$\pm$.011& &\\
    $f_{\Upsilon(1S)}$&0.9005&0.708$\pm$.008&0.645&0.498$\pm$.020& &0.1822\\
    $f_{\Upsilon(2S)}$&0.6072&0.482$\pm$.010&0.439&0.366$\pm$.027& & 0.1944\\
    $f_{\Upsilon(1D)}$&0.5972&&&0.261$\pm$.021& &\\
    $f_{\Upsilon(3S)}$&0.5446&0.346$\pm$.050 &0.393&0.304$\pm$.027& &\\
    $f_{\Upsilon(2D)}$&0.5755&&&0.155$\pm$.011& &\\
    $f_{\Upsilon(4S)}$&0.5859&0.3406$\pm$.00037&&0.259$\pm$.022& &\\
    $f_{\Upsilon(3D)}$&0.6274&&&0.155$\pm$.011& &\\

   \hline
   \end{tabular}
   \end{center}
   \caption{Leptonic decay constants, $\textit{f}_{V}$ of ground state
(1S) and excited state (2S) and (3S) heavy vector quarkonium,
$J/\psi$ and $\Upsilon$ in BSE-CIA along with their masses in other
models and experimental data (all units are in GeV).}
     \end{table}

The calculated values of decay constants are given in Tables 6 and
7. We next calculate the decay widths for two photon and two gluon
decays of pseudoscalar quarkonia $\eta_{c}$ and $\eta_{b}$ and
their radially excited states in the next section.

\section{Two photon and two gluon decays of pseudoscalar quarkonium}

\begin{figure}[h]
\centering
\includegraphics[width=10cm]{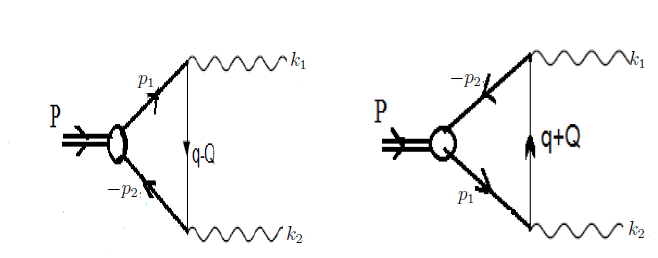}\\
\caption{Diagrams contributing to process $P\rightarrow
\gamma\gamma$. The second diagram is obtained from the first
diagram by reversing the direction of internal fermion lines.}
\end{figure}

The two-photon decays of these states have been the subject of
numerous studies aimed at further understanding the accuracy of
theoretical models of the charmonium and bottomonium systems based
on the available data. The process, $P\rightarrow\gamma\gamma$
proceeds through the famous quark triangle diagrams shown in
Fig.5. In this figure the second diagram is obtained from the
first one by reversing the directions of the internal fermion
lines in the quark loop. Let, $k_{1,2}$, and $\epsilon_{1,2}$ be
the momenta and polarization vectors of the two outgoing photons.
Let, $p_{1}$ and $p_{2}$ be the momenta of constituent quark and
anti-quark constituting the hadron with the total momentum
$P=p_{1}+p_{2}$ and relative momentum $q_{\mu}$. $\Psi(P,q)$ is
the 4D hadron Bethe-Salpeter wave function. For sake of
convenience, we introduce the relative momentum of the two
outgoing photons: $2Q=k_{1}-k_{2}$. In terms of $P$ and $Q$, we
can express the momenta of the outgoing photons as:
$k_{1,2}=\frac{1}{2}P\pm Q$. The momenta of the third quark in the
two diagrams can be expressed as: $p=q-Q$ and $p=q+Q$
respectively.

The amplitude for the process in Fig.5 can be expressed as the sum
of amplitudes for the two diagrams in this figure as:
\begin{equation}
M_{fi}(P\rightarrow\gamma\gamma)=i\sqrt{3}(ie_{q})^{2}\int\frac{d^{4}q}{(2\pi)^{4}}
Tr\{\Psi^{P}(P,q)[\slashed{\epsilon}_{1}S_{F}(q-Q)\slashed{\epsilon}_{2}
 +\slashed{\epsilon}_{2}S_{F}(q+Q)\slashed{\epsilon}_{1}]\}
\end{equation}

$e_{q}=\frac{+2}{3}e$ for charm quark and $e_{q}=\frac{-1}{3}e$
for bottom quark, $S_{F}(q\mp Q)$ are the quark and anti-quark
propagators which are given as:
\begin{equation}
 S_{F}(q\mp Q)=\frac{(\slashed{q}\mp\slashed{Q})\pm m}{(q\mp
 Q)^{2}\mp m^{2}}.
\end{equation}
Now for heavy hadrons like $c\overline{c}$ and $b\overline{b}$,
where the system is basically non-relativistic, it is convenient
to take the internal momentum $q<< M$, and hence, $q^{2}<<Q^{2}$,
where it can be easily seen that $Q^{2}=\frac{M^{2}}{4}$. We now
make use of the fact that the 4D volume element,
$d^{4}q=d^{3}\hat{q}Md\sigma$, and use the relationship between 3D
and 4D BS wave functions. And since the rest of the integrand does
not involve $q$, we can express the amplitude in the above Eq.(53)
as:

\begin{equation}
M_{fi}(P\rightarrow\gamma\gamma)=
\frac{i\sqrt{3}(ie_{q})^{2}}{m^{2}+M^{2}/4}\int\frac{d^{3}\hat{q}}{(2\pi)^{3}}Tr\{\psi^{P}(\hat{q})[\slashed{\epsilon}_{1}
(m+i\slashed{Q})\slashed{\epsilon}_{2}+
\slashed{\epsilon}_{2}(m-i\slashed{Q})\slashed{\epsilon}_{1}]\}.
\end{equation}

Now for a pseudoscalar meson with state $J^{PC}=0^{-+}$, the general
relativistic Salpeter wave function
  $\psi(\hat{q})$
 can be written as:
\begin{equation}
\psi^{P}(\hat{q}) =N_{P}\phi_{P}(\hat{q})
[M+\slashed{P}+\frac{\slashed{\hat{q}}\slashed{P}}{m}]\gamma_{5}
\end{equation}
where $N_{P}$ is the standard BS normalizer which enters into the
BS wave function, and $m$ and M are the masses of quarks and the
corresponding quarkonium respectively. The wave function
$\phi_{P}(\hat{q})$, represents the eigenfunction of the
Pseudoscalar meson obtained by solving the full instantaneous
Bethe-Salpeter equation. For ground state ($1S$) mesons,
$\phi_{P}(\widehat{q})=e^{-\frac{\hat{q}^2}{2\beta^2}}$. The wave
functions for excited states can similarly be expressed, while the
BS normalizer $N_P$ is given as in Eq.(49). Putting the Salpeter
wave function $\psi^{P}(\hat{q})$ into the amplitude in Eq.(55),
and performing the trace over the gamma matrices, we obtain:
\begin{equation}
M_{fi}(P\rightarrow \gamma\gamma)= [F_{P}]
\varepsilon_{\mu\nu\alpha\sigma}P_{\mu}\epsilon_{1\nu}Q_{\alpha}\epsilon_{2\sigma},
\end{equation}

where $F_{P}$ is the radiative decay constant for two-photon decays
of pseudoscalar meson and is expressed as:
\begin{equation}
F_{P}=\frac{8\sqrt{3}e_{q}^{2}}{m^{2}+M^{2}/4}N_{P}\int\frac{d^{3}\hat{q}}{(2\pi)^{3}}\phi_{P}(\hat{q})
\end{equation}
The decay width for $P\rightarrow\gamma\gamma$ is related to the
corresponding decay constant, $F_P$ by the expression:
\begin{equation}
\Gamma_{P\rightarrow\gamma\gamma}=\frac{|F_{P}|^{2}M^{3}}{64\pi}
\end{equation}
The numerical results of $\Gamma_{P\rightarrow\gamma\gamma}$ are
given in the Table 8. We now work out the process of two-gluon
decays of pseudoscalar quarkonia.
\bigskip
\begin{table}[htbp]
   \begin{center}
\begin{tabular}{llllllll}
  \hline
      &BSE-CIA &Expt.\cite{olive14}&BSE\cite{kim}&BSE\cite{munz}&RQM\cite{ebert}&\cite{lansberg}&Pot.Model\cite{bhagyesh11} \\\hline
    $\Gamma_{\eta_{c}(1S)\rightarrow\gamma\gamma}$&7.9178& 7.2$\pm$1.2&7.14 &3.5 &5.5&7.5-10 &11.17 \\
    $\Gamma_{\eta_{c}(2S)\rightarrow\gamma\gamma}$&5.7889&          &4.44 &1.38&1.8&3.5-4.5&8.48  \\
    $\Gamma_{\eta_{c}(3S)\rightarrow\gamma\gamma}$&0.2995&          &     &0.94&   &       &7.57     \\
    $\Gamma_{\eta_{b}(1S)\rightarrow\gamma\gamma}$&0.7376&          &0.384&0.22&0.35&0.560 &0.58  \\
    $\Gamma_{\eta_{b}(2S)\rightarrow\gamma\gamma}$&0.5076&          &0.191&0.11&0.15&0.269 &0.29   \\
    $\Gamma_{\eta_{b}(3S)\rightarrow\gamma\gamma}$&0.4261&          &     &0.084&0.10&0.208&0.24     \\\hline
      \end{tabular}
   \end{center}
   \caption{Two-photon decay widths of ground state (1S) and excited
state (2S) and (3S) pseudoscalar mesons, $\eta_{c}$ and $\eta_{b}$
in present calculation (BSE-CIA) along with their masses in other
models and experimental data (all values are in units of Kev).}
 \end{table}

The two-gluon decay width gives information on the total width of
the corresponding quarkonium. The diagrams for two-gluon decays of
quarkonium can be easily obtained from the diagrams for two photon
decays of pseudoscalar quarkonia (in Fig.5), with a simple
replacement of photons by gluons, and hence, the two quark-photon
vertices with the corresponding quark-gluon vertices. This would
lead to the replacement:
$\alpha\rightarrow\frac{3}{2\sqrt{2}}\alpha_{s}$
 \cite{laverty},
in the expression for $F_{P}$ in Eq.(58), entering into the
two-photon decay width formula, with $\alpha_{s}$ being the QCD
coupling constant. Here we have taken
$e_{q}=+\frac{2}{3}\sqrt{4\pi\alpha}$ for c-quark, and
$e_{q}=-\frac{1}{3}\sqrt{4\pi\alpha}$ for b-quark.
 The results are shown in Table 9.
\begin{table}[htbp]
   \begin{center}
\begin{tabular}{lllllll}
  \hline
     &BSE-CIA &Expt. &BSE\cite{kim} &Pot.Model\cite{bhagyesh11}&BSE\cite{laverty}&Pot. Model\cite{olsson} \\\hline
    $\Gamma_{\eta_{c}(1S)\rightarrow gg}$&13.0699& 26.7$\pm$ 3.0 &19.6 &32.44  &10.57    & 9.010\\
    $\Gamma_{\eta_{c}(2S)\rightarrow gg}$&9.5340& 14.0$\pm$ 7.0  &12.1 &24.64 &5.94 &    \\
    $\Gamma_{\eta_{c}(3S)\rightarrow gg}$&4.4123&              &     &21.99 &     &      \\
    $\Gamma_{\eta_{b}(1S)\rightarrow gg}$&10.8646&              &6.98 &13.72 &12.39&      \\
    $\Gamma_{\eta_{b}(2S)\rightarrow gg}$&7.4766&              &3.47  &6.73 &5.61 &      \\
    $\Gamma_{\eta_{b}(3S)\rightarrow gg}$&6.2763&              &      &5.58 &4.11 &      \\\hline
   \hline
   \end{tabular}
   \end{center}
   \caption{Two-gluon decay widths of ground state (1S) and excited
state (2S) and (3S) pseudoscalar mesons, $\eta_{c}$ and $\eta_{b}$
in present calculation (BSE-CIA) along with their masses in other
models and experimental data (all values are in units of Mev).}
 \end{table}
\section{Numerical Results and Discussions}
We have employed a 3D reduction of BSE (with a $4\times 4$
representation for two-body ($q\overline{q}$) BS amplitude) under
Covariant Instantaneous Ansatz (CIA) for deriving the algebraic
forms of the mass spectral equations whose analytic solutions
(both eigen functions and eigen values), in Eq.(42-43), lead to
mass spectra for ground and excited states of both pseudoscalar
($\eta_{c}$, and $\eta_{b}$) and vector ($J/\Psi$, and $\Upsilon$)
quarkonia, in an approximate harmonic oscillator basis. The
masses, and the eigen functions so obtained are used for
calculating the leptonic decay constants, weak decay constants,
two-photon decay widths, and two-gluon decay widths for ground and
excited states of these pseudoscalar and vector quarkonia.

The mass spectrum calculated in this BSE framework for
($1S$,...,$4S$) states of $\eta_{c}$ and $\eta_{b}$, while for
($1S$, $2S$, $1D$, $3S$, $2D$,...,$5D$) states of $J/\psi$ and
$\Upsilon$ are shown in Table 3 and 4 respectively. All numerical
calculations have been done using Mathematica. We selected the
best set of 6 input parameters, that gave good matching with data
for masses of ground and excited states of $\eta_{c}$, $\eta_{b}$,
$J/\Psi$ and $\Upsilon$ mesons. This input parameter set was found
to be $C_{0}=0.21$, $\omega_{0}$=0.15 GeV., $\Lambda$=0.200 GeV,
and $A_{0}$=0.01, along with the input quark masses $m_c=1.490$
GeV. and $m_b= 5.070$ GeV. The same set of parameters above was
used to calculate the leptonic decay constants of $\eta_{c},
\eta_{b}, J/\psi$, and $\Upsilon$, as well as the two-photon and
two-gluon decay widths of $\eta_{c}$, and $\eta_{b}$. However the
experimental data on masses and decay constants/ decay widths of
many of these states is not yet currently available. The results
obtained for masses of ground and radially excited states of
$\eta_{c}$, $\eta_{b}$, shown in Table 3, are in reasonable
agreement with experiment. However, a wide range of variation in
masses of various states in different models such as Lattice QCD
model \cite{burch09}, and QCD sum rule model \cite{veli12} can be
seen in Table 3.

As regards the mass spectral predictions for vector mesons is
concerned, we have listed the values for $M_{J/\psi}$ for $1S$,
$2S$, $1D$,...$5S$, and for $M_{\Upsilon}$ for $1S$, $2S$,
$1D$,...,$5D$ in Table 4. Many other states such as $3D$, $4D$,
and $5D$ for $J/\Psi$ are not yet experimentally available. The
same holds true for all the $D$ states of $\Upsilon$. For vector
$c\overline{c}$ quarkonia, the masses of ground and excited states
are very close to central values of data. However the $1S$, and
$2S$ states of $b\overline{b}$ are somewhat over estimated from
central values of data. The disagreement with data increases as we
go to the higher excited states of $b\overline{b}$. This is due to
the fact that we did not incorporate the one-gluon-exchange (OGE)
effects in the kernel, and used only the confining part of
interaction as in our Eq.(31) (taking analogy from
\cite{koll,babutsidze,hluf15,olsson} for heavy quarkonia). Our
results reflect the fact that the OGE term becomes more and more
important as we go to very heavy $b\overline{b}$ quarkonia, where
the distance between the quark and the anti-quark will be tiny,
and its contribution to their mass spectra will be substantial.
Further, there is a degeneracy in the masses of $S$ and $D$ states
with the same principal quantum number $N$ for $J/\Psi$ and
$\Upsilon$.  The inclusion of OGE terms in the potential will also
lift up the degeneracy in these states and also bring masses of
$b\overline{b}$ vector quarkonia closer to experiment.

However, we wish to mention that in this paper, our main emphasis
was to show that this problem of $4\times 4$ BSE under heavy quark
approximation can indeed be handled analytically for both masses,
as well as the wave functions in an approximate harmonic
oscillator basis. The  analytical forms of wave functions obtained
as solutions of mass spectral equations (that are derived from 3D
BSE), were then used to calculate the decay constants and decay
widths for various processes involving these quarkonia. We next
intend to incorporate the OGE effects perturbatively into our mass
spectral equations, using our wave functions so obtained in
Eqs.(41) as unperturbed wave functions in this study.

As regards our results for decay constants is concerned, it is
seen that our calculated $f_P$ value of ground state is
$f_{\eta_c(1S)}=0.4044$GeV., and is within the error bars of its
experimental value, $0.335\pm 0.075$Gev \cite{cleo01}. However a
wide range of predictions of decay constant values for all states
can be seen in different models in Table 6. The experimental data
for many of these states is not yet available. However, it is
observed that the decay constants keep decreasing as one goes from
$(1S)$ to $(3S)$ states for both $\eta_c$ and $\eta_b$ mesons.
This also signifies that the instability of these states increases
with increase in radial quantum number $N$. This trend is similar
to the trend observed in a recent potential model calculation
\cite{bhagyesh11} of decay constants of $\eta_c$ and $\eta_b$.

Our results for leptonic decay constants of ground and excited
states of $J/\psi$ and $\Upsilon$ are listed in Table 7. Our decay
constants for $J/\psi$ and $\Upsilon$ for excited states are
somewhat on the higher side in comparison to central values of
data for these states. However, the decay constant $f_{V}$ values
of various models again show a very wide range of variation as can
be seen from Table 7.

Then, with the same set of parameters, we calculate the two-photon
radiative decay widths of the ground and first two radially
excited states of $\eta_{c}$ and $\eta_{b}$ mesons. The two photon
decay widths of our model are listed in Table 8. Our result for
two-photon decay width of ground state of $\eta_c$ is
$\Gamma_{\eta_{c}}(1S)=7.9178$KeV., and is within the error bars
of data, $\Gamma_{\eta_{c}(1S)}(Exp.)=7.2\pm 1.2$ KeV
\cite{olive14}. We have compared our results with those of other
models, though data on many of these is not yet available. The two
photon decay widths of other models also show a wide range of
variation.

We then calculate the two-gluon decay widths in our model for
ground and radially excited states of $\eta_{c}$ and $\eta_{b}$,
which are shown in Table 9. The two-gluon decay process accounts
for a substantial portion of hadronic decay widths for states
below $c\bar{c}$ or $b\bar{b}$ threshold. However, as pointed out
in \cite{laverty}, due to significant contributions from radiative
corrections as well as from three-gluon decays, the two gluon mode
does not give the complete picture. Our results on two-gluon
widths are thus smaller than the hadronic widths of $\eta_{c}$ and
$\eta_{b}$ states. However as can be seen from Table 9, the
results of two-gluon decay widths in various models again show a
wide range of variations.

However, as mentioned earlier, our main emphasis in this paper was
to show that this problem of $4\times 4$ BSE under heavy quark
approximation can indeed be handled analytically for both masses,
as well as the wave functions for $l=0$, and $l=2$ states. The
validity of heavy quark approximation for quarkonia is due to the
fact that not only the relative momentum between heavy quarks in
the bound states is considered small, but also these quarks are
treated as almost on mass shell \cite{syli14}, which is justified
for calculation of low energy properties like the mass spectrum,
and the decays of quarkonia ($c\bar{c}$ and $b\bar{b}$ systems).
With the above approximation, that is quite justifiable and well
under control, in the context of heavy quark systems, we have been
able to give analytical solutions of mass spectral equations
Eqs.(35-36) of both pseudoscalar and vector quarkonia, giving us a
much deeper insight into this problem.

Analytical forms of the wave functions for both $S$ and $D$ states
for $n=0,1,2,...$ for $c\overline{c}$, and $b\overline{b}$ systems
thus obtained are given in Eq.(41). These wave functions  were
then used to calculate the decay constants and decay widths for
various processes involving these quarkonia. We have also plotted
the graphs of all the wave functions for the states, $1S,...,4S$
for $\eta_c$ and $\eta_b$ mesons, and for the states
$1S,2S,1D,3S,2D,4S$, and $3D$ for $J/\Psi$ and $\Upsilon$ mesons,
in Figs. 1-4. The over all features of all these plots show that
the states $nS$ and $nD$ have $n-1$ nodes in their wave functions.

We are not aware of any other BSE framework, involving $4\times 4$
BS amplitude, that treats the mass spectral problem involving
heavy quarkonia analytically. To the best of our knowledge, all
the other $4\times4$ BSE approaches treat this problem numerically
just after they obtain the coupled set of Salpeter equations (see
\cite{wang10}). We further wish to mention that the over all
features of our plots of wave functions in Eqs.(41) derived in our
framework are very similar to the corresponding plots of wave
functions obtained by purely numerical methods in \cite{wang10},
suggesting that our approach is not only in good agreement with
the numerical approaches followed in other works, but also gives a
deeper insight into the problem, by showing an explicit dependence
of the mass spectrum on principal quantum number $N$ as in
Eqs.(42-43). As mentioned above, we next intend to incorporate the
OGE effects into the mass spectral equations perturbatively.
However, this we intend to do as further work.

We also intend to extend this study to calculation of observables
of heavy-light mesons such as $D$ and $B$, which would involve
incorporation of unequal mass kinematics in our framework, which
is beyond the scope of the present paper, and we wish to do as
further work. This study will also be extended to processes
involving quark-triangle diagrams with two or more hadronic
vertices such as to decays $V\rightarrow P\gamma$, and
$V\rightarrow PP$ (with $P$ and $V$ being the pseudoscalar and
vector quarkonia respectively), as further works. It is further
expected that our present framework (with $4\times 4$
representation of two-body BS amplitude) will be able to overcome
the complexities in the amplitudes for these processes that
appeared in our earlier
framework\cite{bhatnagar13}.\\

\textbf{Acknowledgements}: This work was carried out at Addis Ababa
University (AAU). We are thankful to the Physics Department, AAU for
the facilities provided during the course of this work. HN thanks
Samara University for support for his Doctoral programme.


\end{document}